\documentclass[12pt]{article}

\usepackage[mathscr]{euscript}
\usepackage[makeroom]{cancel}
\usepackage[font={small,it,singlespacing}]{caption}
\usepackage{amsmath, amssymb, amsthm, amsbsy, amsfonts, bigints}
\usepackage{hyperref, enumerate,enumitem, verbatim, sectsty, colortbl, tabularx}
\usepackage{natbib}
\usepackage{graphicx, float, pdflscape,rotating, pspicture, epsfig, subcaption, threeparttable, multirow, multicol, afterpage,lscape}
\usepackage[compact]{titlesec}
\usepackage[table]{xcolor}
\usepackage{longtable, float}
\usepackage[skins]{tcolorbox}
\usepackage{standalone}
\usepackage[belowskip=2pt,aboveskip=2pt]{caption}
\usepackage[bottom]{footmisc}
\usepackage[nodisplayskipstretch]{setspace}
\usepackage{hyperref}

\hypersetup{colorlinks,citecolor=blue}
\allowdisplaybreaks

\renewcommand{\qedsymbol}{$\blacksquare$}

\newcommand{\R}{\mathcal{R}}

\newcommand{\x}{\mathbf{x}}
\newcommand{\s}{\mathbf{s}}

\newcommand{\var}{\mbox{var}}
\newcommand{\cov}{\mbox{cov}}

\newcommand{\E}{\mbox{E}}
\newcommand{\V}{\mbox{V}}

\theoremstyle{plain}

\newtheorem{defn}{Definition}

\newtheorem{pro}{Proposition}
\newtheorem{rem}{Remark}

\newcommand{\maxx}{\underset{
\begin{subarray}{c}
  \scriptstyle{\mathbf{x},\mathbf{x}'} \\
  \scriptstyle{\Delta(\mathbf{x},\mathbf{x}')=1}
 \end{subarray}}{\mbox{max}}}

\newcommand{\maxxu}{\underset{
\begin{subarray}{c}
  \scriptstyle{\mathbf{x},\mathbf{x}',\s^*\in\mathcal{S}} \\
  \scriptstyle{\Delta(\mathbf{x},\mathbf{x}')=1}
  \end{subarray}}{\mbox{max}}}

\graphicspath{{codes/}}

\begin{document}
\title{\large{\textbf{Statistical Properties of Sanitized Results from  Differentially Private  Laplace Mechanism with Univariate Bounding Constraints}}}
\author{\small{\textbf{ Fang Liu\footnote{Fang Liu is Associate Professor in the Department of Applied and Computational Mathematics and Statistics, University of Notre Dame, Notre Dame, IN 46556 ($^{\ddag}$E-mail: fang.liu.131@nd.edu). The work was supported by the NSF Grants 1546373, 1717417, and the University of Notre Dame Faculty Research Initiation Grant}}}\\
\small{Department of Applied and Computational Mathematics and Statistics} \\
\small{University of Notre Dame, Notre Dame, IN 46556}\\
\date{}}
\maketitle
\vspace{-2cm}
\begin{abstract}
\noindent
Protection of individual privacy is a common concern when releasing and sharing data and information. Differential privacy (DP)  formalizes privacy in probabilistic terms without making  assumptions about the background knowledge of data intruders, and thus provides a robust concept for privacy protection.  Practical applications of DP involve development of differentially private mechanisms to generate sanitized  results at a pre-specified privacy budget. For the sanitization of  statistics with publicly known bounds such as proportions and correlation coefficients, the bounding constraints will need to be incorporated in the differentially private mechanisms. There has been little work on examining the consequences of the bounding constraints  on the accuracy of sanitized results and the statistical inferences of the population parameters based on the sanitized results.   In this paper, we  formalize the  differentially private truncated and boundary inflated truncated (BIT)  procedures  for releasing statistics with publicly known bounding constraints. The impacts of the  truncated and BIT Laplace procedures on the statistical accuracy and validity of sanitized statistics are evaluated both theoretically and empirically via simulation studies. 

\vspace{0.5cm}
\noindent  \textit{\textbf{keywords}}:  truncated  mechanism, boundary inflated truncated (BIT) mechanism, bias, consistency, mean squared error; global and data-invariant

\end{abstract}

\newpage
\section{Introduction}\label{sec:introduction}
Protection of individual privacy is always a concern when releasing and sharing information. A data release mechanism aims to provide useful information to the public without compromising individual privacy.  Differential privacy (DP) is a concept  developed in theoretical computer science \citep{dwork2006calibrating, dwork2008, dwork2011differential} and has gained great popularity in recent years in both theoretical research and practical applications.  DP formalizes privacy in mathematical terms without making  assumptions about the background knowledge of data intruders and thus provides a robust concept for privacy protection.  Practical applications of DP involve development of differentially private mechanisms, also referred to as sanitizers, through which original results are processed and converted to results that do not reveal individual information at a pre-specified privacy budget.    There are general differentially private mechanisms such as the Laplace mechanism \citep{dwork2006calibrating}, the Exponential mechanism \citep{mcsherry2007mechanism, mcsherry2009privacy}, and more recently, the staircase mechanism \citep{staircase}, the generalized Gaussian mechanism \citep{liu2016a},  and the adaptive mechanisms such as  the multiplicative weighting mechanism \citep{multiplicative} and the median mechanism \citep{roth2010median} for sanitizing multiple correlated queries. There are also differentially private mechanisms targeting specifically at certain statistical  analyses such as robust and efficient point estimators \citep{dwork2010differential, dwork2011differential}, principle component analysis \citep{Chaudhuri2012PCA}, linear and penalized regression \citep{Chaudhuri2011, Kifer2012}, Bayesian inferences of probabilistic graphical models \citep{zhang},  machine learning, data mining,  and big data analytics in genomics, healthcare, biometrics \citep{blum2008learning, mohammed2011differentially, Yu2014,connectedhealth, biometrics}, among others.

In the context of DP, it is sometimes assumed that data and statistics (numerical query results) are bounded. The bounding assumption is supported from a technical perspective as well as justified from a practical point of view.  First,  some statistics are  naturally bounded per definition, such as proportions  (bounded by $[0,1]$) and correlation coefficients (bounded by $[-1,1]$). Second, some statistics are required be bounded in order to be sanitized by some of the common differentially private mechanisms while ensuring some degree of usefulness of the sanitized results. For example, the scale parameter of the Laplace mechanism is proportional to the global sensitivity of a statistic. Suppose the statistic is the sample mean of a numerical attribute; then the value of the attribute needs to be bounded on both ends to have a finite global sensitivity for the mean so that it can be sanitized by the Laplace mechanism in a meaningful way. Third, real-life data in general  support the assumption of bounded data when the assumption is needed. 
Though bounded numerical attributes in statistical parametric modelling are often modelled via distributions with unbounded domains (e.g., Gaussian or Poisson assumptions), those distributional assumptions are  in many cases  only approximate and the probabilities of out-of-bounds values are often small enough to be ignorable under these distributional assumptions.  For instance,  it is safe to say human height is bounded within $(0, 300)$cm. Though it is often modelled by  Gaussian distributions with support $[-\infty,\infty]$,  $\Pr$(height $<0$ cm or $>300$ cm) $\approx 0$ under the Gaussian assumption.  If all numerical attributes in a data set are bounded, descriptive or inferential statistics based on the data in general  are  also bounded. For example, if a numerical attribute is bounded within $[c_0,c_1]$, then its sample mean is bounded within $[c_0,c_1]$, and its variance is bounded within $[0, n(c_1-c_0)^2/(4(n-1))]$  for a given sample size $n$ \citep{shiffler1980upper}. Besides the above examples on the univariate bounding constraint per statistic, there are also many types of multivariable bounding constraints. For example, a linear combination of multiple statistics is bounded or the sum of a proportion vector is  1. In the following discussion, we focus our theoretical and empirical analysis on the univariate bounding and discuss a couple examples of multivariable constraints briefly; more in-depth investigation on the latter topic will be conducted in the future.

Modifications are often needed in some  commonly used differentially private mechanisms in order to accommodate the release of bounded statistics  without compromising the pre-specified DP. For example, both the Laplace and Gaussian mechanisms release sanitized results from the real line ($-\infty, \infty$) and do not automatically deal with bounding constraints. Some practitioners choose to ignore the bounds and release the raw sanitized results as is. We would not recommend this approach  since the out-of-bounds values carry no practical meaning and eventually will be discarded by data users.  Another approach is to formulate the problem as a constrained optimization problem with inequality constraints  in the general framework of  constrained inference in DP. The constrained inference approach concerns about finding a set of sanitized results that are optimal estimators of the unconstrained sanitized results by some criteria, such as the $l_2$ distance, subject to a set of predefined and known constraints. Therefore, ``inferences'' in this context is not the same as the classical statistical inferences, which is about inferring population parameters via point and interval estimations and hypothesis testing given finite sample data. The constrained inference can deal with   both inequality and equality constraints. For instance, proportions $\mathbf{p}=(p_1,\ldots,p_K)^T$ are subject to both the equality  constraint $\sum_{j=1}^K p_j=1$ and the inequality constraint  $p_j\in[0,1]$ for $j=1,\ldots,K$. \citet{barak2007privacy} employed linear programming (and the Fourier transformation) to obtain a non-negative and consistent sanitized contingency tables. \citet{boosting} boosted the accuracy of sanitized histograms (measured by the mean squared error  between the sanitized and original results) by incorporating the prior rank constraint in unattributed histograms  and the equality constrained in universal histograms to increase the accuracy of lower-order marginals. \citet{hierarchical} showed that combining the choice of a good branching factor with constrained inference can further boost the accuracy of a sanitized histogram.  \citet{matrix} investigated an extension to the matrix mechanism they proposed that incorporates non-negativity constraints when realizing count queries. While the constrained inference approach provides a general framework to deal with constraints in DP,  all the above mentioned work deals with count data. In addition, the approach can be analytically and computationally demanding, depending on the statistics and the objective functions employed. We will save the in-depth investigation of this approach as a  future research topic.

In this paper, we focus on examining two  straightforward methods that appeal to practitioners for sanitizing statistics with bounding constraints. One approach, referred to as the \emph{truncation} procedure, throws the out-of-bounds sanitized values  away and re-sanitizes until a within-bound value is obtained. This procedure can also be realized by sampling sanitized results directly from a truncated distribution that satisfies DP.   The other approach legitimizes  the out-of-bounds sanitized  results by setting the out-of-bounds values at the boundaries, referred as the \emph{boundary inflated truncation/BIT} procedure. This procedure can also be realized by sampling sanitized results directly from a piece-wise distribution with probability mass at the boundaries. We  assess  their impact  on the statistical accuracy of sanitized results vs the original results both theoretically and empirically in the context of the Laplace mechanism. To the best of our knowledge, this is the first work on the statistical properties of the BIT and truncation bounding procedures in the context of DP. 

The rest of the paper is organized as follows. Section \ref{sec:prelim} overviews some of the key concepts in DP. Section \ref{sec:method} presents the truncation and BIT bounding procedures and investigates the impact of the BIT and truncation bounding procedures on the utility of sanitized results in terms of bias, mean squared error, and consistency.  Section  \ref{sec:simulation} illustrates the applications of the truncated and BIT Laplace procedures and examines the statistical properties of the sanitized results in two  simulation studies. The paper concludes in Section \ref{sec:discussion} with some final remarks and plans for future works.

\section{Preliminaries}\label{sec:prelim}
\subsection{Definition of Differential Privacy}
Let $\Delta(\x,\x')=1$ denote that data $\x'$ differs from $\x$ by only one individual. There are two  definitions on ``differing by one''.  Briefly, one refers to the case that the $\x'$ and $\x$  have the same sample size $n$, but one and only one record differs in at least one attributes (a substitution would make the two data sets identical). In the second definition, $\x'$ has one less record that $\x$, so the sample sizes for $\x$ is $n$ and $n-1$ for $\x'$) (a deletion or a insertion would make the two data sets identical)  $\epsilon$-differential privacy (DP) is defined as follows.
\begin{defn}[\textbf{$\epsilon$-Differential privacy} \citep{dwork2006calibrating}]\label{defn:dp} A randomized mechanism $\mathcal{R}$ satisfies $\epsilon$-differential privacy if for all  all data sets $(\x,\x')$ that is $\Delta(\x,\x')=1$ and all result subsets $Q$ to query $\mathbf{q}$,  $\left|\log\left(\frac{\Pr(\R(q(\x))) \in Q)}{\Pr(\R(q(\x'))\in Q)} \right)\right|\le\epsilon$, where $\epsilon>0$ is the privacy budget parameter.
\end{defn}
Under DP,  the ratio in the probabilities of  obtaining the same query results from $\x$ and $\x'$ after sanitization via $\R$ is bounded within $(e^{-\epsilon}, e^{\epsilon})$ -- a neighborhood around 1. For a small $\epsilon$ value, the ratio is close to 1, meaning that the query result is similar with or without that individual in the data set, and the chance that a participant in the data will be identified based on the query result sanitized  via $\R$ would be low. DP provides a robust and powerful model against privacy attacks in the sense that it does not make  assumptions on the background knowledge or the behavior on data intruders. $\epsilon$ can be used as  a tuning parameter -- the smaller $\epsilon$ is, the more protection there is on the released information via $\R$. 

There also exist softer versions to the pure $\epsilon$-DP that include the  $(\epsilon,\delta)$-approximate differential privacy (aDP) \citep{dwork2006delta}, the $(\epsilon,\delta)$-probabilistic DP (pDP) \citep{onthemap}, the $(\epsilon,\delta)$ random differential privacy (rDP) \citep{randomDP}, and the $(\epsilon,\tau)$-concentrated differential privacy (cDP) \citep{cPD}.  In all the relaxed versions,  one additional parameter is employed  to characterize the amount of relaxation on top of the privacy budget $\epsilon$.   In $(\epsilon,\delta)$-aDP,  $\Pr(\R( q(\x)) \in Q)\le e^{\epsilon}\Pr(\R( q(\x))\in Q) + \delta$. A sanitization algorithm satisfies $(\epsilon,\delta)$-probabilistic differential privacy (pDP) if the probability of generating an output belonging to the disclosure set is bounded below $\delta$, where the disclosure set contains all the possible outputs that leak information for a given privacy tolerance $\epsilon$. The $(\epsilon,\delta)$-rDP is also a probabilistic relaxation of DP; but it differs from $(\epsilon,\delta)$-pDP in that the probabilistic  relaxation  is with respect to the generation of the data  while  it is with respect to the sanitizer in the  $(\epsilon,\delta)$-pDP.  The $(\epsilon,\tau)$-cDP, similar to the $(\epsilon,\delta)$-pDP, relaxes the satisfaction of DP with respect to the sanitizer, and ensures that the expected privacy cost is $\epsilon$ and (Prob(the actual cost $>\epsilon$)$>a$)  is bounded by $e^{-(a/\tau)^2/2}$.

\subsection{Laplace Mechanism and Global Sensitivity}
The Laplace mechanism   is  a popular sanitizer to release statistics with $\epsilon$-DP   \citep{dwork2006calibrating}. \citet{liu2016a} introduces the generalized Gaussian mechanism $(\epsilon,\delta)$-pDP that includes  the Laplace mechanism  as a special case (when $p= 1$ and $\delta=0$). Denote the statistics of interest by $\s_{r\times1}$. The Laplace and the generalized Gaussian mechanisms are based on the $l_p$ global sensitivity, which is defined as
\begin{equation}
\delta_p=\maxx\|\s(\x)-\s(\x')\|_p=\maxx\!(\textstyle\sum_{j=1}^r|s_j(\x)-s_j(\x')|^p)^{1/p}
\end{equation}
for all pairs of data  sets $(\x,\x')$ that are $\Delta(\x,\x')=1$.  $\delta_p$  is the maximum difference in $\s$ in the $l_p$ distance between any pair of data sets $\x,\x'$ with  $\Delta(\x,\x')=1$. The sensitivity is ``global" since it is defined for all possible data sets and all possible ways of two data sets differing by one record. The larger the global sensitivity for $\s$, the larger the disclosure risk is from releasing the original $\s$, and the more perturbation is needed for $\s$ to offset the risk.  Specifically,  the Laplace mechanism of $\epsilon$-DP sanitizes $\s_{r\times1}$ as in 
\begin{align}\label{eqn:laplace}
\s_{r\times1}^*=(s_1,\ldots,s_r)\textstyle \sim\prod_{j=1}^r \mbox{Lap}\left(s_j,\delta_1\epsilon^{-1}\right),
\end{align}
where $\delta_1=\sum_{j=1}^r\delta_{j1}$ is the $l_1$-global sensitivity of $\s$ and $\delta_{j1}$ is the global sensitivity of the $j$-th statistic in $\s$ . For integer $p\ge2$,  the generalized Gaussian   mechanism of order $p$ sanitizes $\s$ with $(\epsilon,\delta)$-pDP by drawing sanitized $\s^*$ from the generalized Gaussian  distribution 
\begin{equation}
f(\s^*)\!=\!\textstyle\prod_{j=1}^r\! \frac{p}{2b\Gamma(p^{-1})}\exp\{\!(|s^*_j-s_j|/b)^p\} = \prod_{j=1}^r \mbox{GG}(s_j,b,p),
\end{equation}
where $b$ satisfies $\Pr\left(\textstyle\sum_{k=1}^r a_j\! >\!b^p-\epsilon\delta_p^p\right) \!< \!\delta$ with $a_j=\textstyle\sum_{j=1}^{p-1}(_j^p)|s^*_j-s_j|^{p-i}\delta_{1,k}^j$, $\delta_{1,k}=$  the $l_1$-global sensitivity of $s_j$, and $\delta_p=$  the $l_p$-global sensitivity of $\s$. When $p=2$, the generalized Gaussian   mechanism becomes the Gaussian mechanism of  $(\epsilon,\delta)$-pDP  that generates sanitized $s^*_j$ from  N$(s_j,\sigma^2=b^2/2)$ for $k=1,\ldots,r$.  The Laplace  and generalized Gaussian  mechanisms produce  unbound sanitized results from the real line ($-\infty, \infty$); therefore, some bounding procedures will be needed if the mechanisms are to be applied to sanitize bounded statistics.  

The global sensitivity for a query or statistics in general needs to be determined analytically though its value might not be tight. Numerical computation of global sensitivity is not feasible as it is impossible to enumerate all possible data $\x$ and all possible ways of $\Delta(\x,\x')=1$ especially when $\x$ contains continuous attributes or when sample size $n$ is large.  We have obtained the $l_1$ global sensitivity of some common statistics, including proportions, means, variances, and covariances (refer to the online supplementary materials). The global sensitivity values are calculated for both definitions of two data sets differing by one record and the global sensitivity is the same on most of the examined statistics regardless of which definition is used. For example, the global sensitivity of the sample mean of a variable the value of which is bounded within $[c_0,c_1]$ is $n^{-1}(c_1-c_0)$, and that of the sample variance is $n^{-1}(c_1-c_0)^2$. 
In all calculations, we assume the sample size $n$ of the original data is a known and carries no privacy concern, which is often the case in statistical analysis. 
It should be noted that the global sensitivity  of a function of a statistic $s$ is not equal to the function of the global sensitivity of $s$ in general. For example, $\delta_1$ of a sample variance is $(c_1-c_0)^2n^{-1}$, but $\delta_1$ of the sample standard deviation cannot be simply calculated as $\sqrt{(c_1-c_0)^2n^{-1}}$. In fact, the global sensitivity of the standard deviation is more difficult to calculate analytically compared to that of the variance.  When the global sensitivity of $s$ is not easy to calculate, but a data-independent function of $s$, say $t=f(s)$, is, we can instead sanitize $t$ to obtain $t^*$ and then obtain sanitized $s^*$ via the back-transformation $s^*= f^{-1}(t^*)$. 
\section{Truncated and BIT Laplace Mechanisms}\label{sec:method}
In this section, we first formalize  two commonly used bounding procedures and then examine the statistical properties of  sanitized outcomes processed by the twp procedures, respectively. Both procedures are intuitive and straightforward to apply and can be coupled with any  differentially private mechanisms. We focus on their applications in the context of the Laplace mechanism given its popularity, and will look into their applications in other mechanisms as the generalized Gaussian mechanism in the future.

\subsection{Definitions}\label{sec:bounding}
\begin{defn}[\textbf{truncated and boundary-inflated-truncated Laplace mechanisms}]  \label{defn:posthoc}
Denote the bounded statistics  by $\s_{r\times1}=(s_1,\ldots,s_r)\in[c_{10},c_{11}]\times\cdots\times[c_{r0},c_{r1}]$, where $[c_{j0},c_{j1}]$ are the bounds for $i\textsuperscript{th}$ element in $\s$ ($i=1,\ldots,r$), the privacy budget by $\epsilon$, the $l_1$-global sensitivity of $\s$ by $\delta_1$. Let $\lambda= \delta_1\epsilon^{-1}$.
\begin{itemize}[leftmargin=0.25in]
\item[(a)] The \textbf{truncated Laplace mechanism}  of $\epsilon$-DP sanitizes $\s$  by drawing $\s^*$ from  the truncated Laplace distribution
\begin{align}\label{eqn:truncated}
f(\s^*)\!=\!\prod_{j=1}^r\mbox{Lap}(s_j,\lambda|c_{j0}\le\! s^*\!\le c_{j1})
\!=\!\prod_{i=1}^r\frac{\exp\left(-\frac{|s_j^*-s_j|}{\lambda}\right)}{2\lambda\!\left(1-\frac{1}{2}\exp(-\frac{c_{j1}-s_j}{\lambda})\!-\!\frac{1}{2}\exp(\frac{c_{j0}-s_j}{\lambda})\right)}.
\end{align}
\item[(b)] The \textbf{boundary-inflated-truncated (BIT) Laplace mechanism} of $\epsilon$-DP sanitizes  $\s$ by drawing $\s^*$ from the BIT Laplace distribution $f(\s^*)=\prod_{i=1}^rf(s_j^*)$, where $f(s_j^*)$ is
\begin{equation}\label{eqn:BIT}
f(s_j^*) =  \begin{cases}
\frac{1}{2}\exp(-(s_j-c_{j0})/\lambda)  & \mbox{if $s_j^*=c_{j0}$}\\
\mbox{Lap}(s_j, \lambda) & \mbox{if $c_{j0}< s_j^*< c_{j1}$}\\
1-\frac{1}{2}\exp(-(c_{j1}-s_j)/\lambda)  & \mbox{if $s_j^*=c_{j1}$}\\
\end{cases}
\end{equation}
\end{itemize}
\end{defn}
\noindent Rather than sampling directly from Eqn (\ref{eqn:truncated}), the truncated Laplace mechanism can also be realized via in a post-hoc manner by throwing away out-of-bounds  sanitized results from the regular Laplace mechanism until catching an in-bound value.  Similarly, the BIT bounding procedure can be realized by post-hoc  setting out-of-bounds sanitized results   from the regular Laplace mechanism at the corresponding boundaries. 
If  the scale parameter $\lambda\rightarrow\infty$ in the Laplace distribution when $\epsilon\rightarrow 0$ or $\delta_1\rightarrow \infty$, it can be easily established that $f(s_j)$ in the truncated Laplace mechanism in Eq.~(\ref{eqn:truncated}) converges to an uniform distribution unif$(c_{0i},c_{1i})$, and that in the BIT Laplace distribution in Eq.~(\ref{eqn:BIT}) converges to a Bernoulli distribution with probability mass at $c_{0i}$ and $c_{1i}$, respectively. In both of the asymptotic cases, the sanitized results preserve little original information. Both the Laplace and BIT Laplace mechanisms, as in the regular Laplace mechanism, require calculation of the $l_1$-global sensitivity for the targeted $\s$ to be sanitized. 

\begin{rem}[\textbf{multivariable-function constraints}]\label{rem:multi}
Definition \ref{defn:posthoc} focuses on the case where each statistic $s_j$ in $\s$ has its own lower and upper bounds $[c_{0j}, c_{1j}]$ ($j=1,\ldots,r$) that are publicly known and fixed constants. In some practical cases, besides the bounding constraints for each statistic, a subset of $\s$ or all of its elements are also required to satisfy multivariable constraints. For example, $\s$ is a set of proportions under the linear constraint $\sum_{j=1}^rs_j=1$; or a subset $\mathcal{S}$ of $\s$ satisfies $c_0\le \sum_{k\in\mathcal{S}}a_k s_k\le c_1$. In the presence of the additional multivariable constraints, one would first check whether the multivariable constraints are compatible with the bounding constraint on each $s_j$, which may lead to a set of new constraints that incorporate all the constraints -- univariate or multivariate. For example, say $\s=(s_1,s_2)$ and the linear constraint is $s_1+s_2=c$. The final constraints on $s_1$ and  $s_2$ would then be $[\max(c-c_{12},c_{01}), \min(c-c_{02},c_{11})]$ and $[\max(c-c_{11},c_{02}), \min(c-c_{01},c_{12})]$, respectively. After applying either the truncated or the BIT Laplace mechanism in Definition \ref{defn:posthoc} to sanitize, say $s_1$ to obtain $s_1^*$, that satisfies the new bounding constraint, the other can be calculated as $s_2^*=c-s_1^*$. In some special cases, such as when $\s$ is a set of proportions under the linear constraint $\sum_{j=1}^rs_j=1$, one can also apply a simple rescaling procedure to the sanitized  $\s$ to satisfy the multivariable linear constraint (see simulation study  2 in Section \ref{sec:simulation} for an example).
\end{rem}

There are many types of multivariable constraints. Depending on the nature of the additional multivariable constraints, there may exist different approaches to  satisfy all the constraints simultaneously -- some might lead to better original information preservation than others for the same privacy budget. In addition, the computational complexity to obtain a set of sanitized $\s^*$ that satisfy all the constraints may reach a level beyond practicality. The diversity and complexity of  satisfying multivariable constraints also indicates challenges for the theoretical analysis on the statistical properties of sanitized $\s^*$ in this setting.  For these reasons, in the theoretical analysis in Section \ref{sec:property}, we focus on learning the statistical properties of sanitized $\s^*$ where each of its element has only the univariate bounding constraint with no additional multivariable constraints. Incorporating  multivariable-function constraints on top of the univariate bounding constraints deserves more in-depth investigation and will be a topic of future research. 

\subsection{Statistical Properties of  Sanitized \texorpdfstring{$\s^*$ under Univariate Bounding Constraint}{} }\label{sec:property} 
In Definition \ref{defn:posthoc}, the bounds $[c_{10},c_{11}]\times\cdots[c_{r0},c_{r1}] $ are assumed to be data invariant and ``global'' to guarantee $\epsilon$-DP, from the data release perspective. If the bounds are ``local'' and data-specific, meaning they are functions of data  $\x$ at hand, then the bounds themselves, would need to be sanitized and would otherwise leak information about the original data.  On the other hand,  ignoring the local properties of data $\x$ in a bounding procedure could have a negative impact on the statistical properties of sanitized results $\s^*$. In this section, we investigate the statistical behaviors of $\s^*$ produced by a sanitizer with bounding constraints. 

The rationale for studying the statistical  properties of the sanitized statics is as follows. First, duo to randomness of the injected noise to satisfy DP, the sanitized statistic is a random variable (even without accounting for the sampling variability of the sample data after conditional on the original static). Though the privacy budget is often only enough for querying one or a few statistics, the users would still be interested in understanding and gaining theoretical insights on the statistical properties of the sanitized statistic in expectation over the distribution of the injected noise, such as bias and MSE of the sanitized statistic. Furthermore, one of the main goals for data collection is to draw inferences about the population from which the sample data come from. For non-private inferences, one would often examine the statistical properties of an estimator given sample data to see how good it is for a population parameter, such as unbiasedness and consistency. In the context of private inferences, the users won't have access to the original non-private estimate, but rather its sanitized version.  If the sanitized statistic does not possess desirable statistical properties, say it is biased,  for the  original statistic, and even if the original statistic is unbiased  for the population parameter,  then users know the statistical inference based on the sanitized statistic would be biased as well, per the law of total expectation.

We  start by defining the desired statistical properties of sanitized $\s^*$ (Definition \ref{defn:unbiased}), then examine the bias of $\s^*$ sanitized by the  truncated  and BIT Laplace mechanisms relative to the original $\s$ (Proposition \ref{thm:mu12}), and present an upper bound for the mean squared error and examine the consistency of $\s^*$ for $\s$ and the convergence rate  in Proposition \ref{lem:error}. In addition, we list in Proposition \ref{prop:asymp} a set of sufficient conditions for the sanitized $\s^*$ to be asymptotically unbiased and consistent  for the population parameter $\boldsymbol{\theta}$ in the case where the original $\s$ is an estimator  for $\boldsymbol{\theta}$.

\begin{defn}[\textbf{unbiasedness and consistency of sanitized statistic for original statistic}]\label{defn:unbiased}
Sanitized $\s^*$ is unbiased for the original $\s$ if  $\E_{\s^*}(\s^*|\s)= \s$;  $\s^*$ is asymptotically unbiased for  $\s$ if  $\E_{\s^*}(\s^*|\s)\rightarrow \s$ as $n\rightarrow\infty$, where $n$ is the sample size of original data $\x$;  $\s^*$ is consistent for  $\s$ if  $\s^*\xrightarrow{p}{} \s$ as $n\rightarrow\infty$.
 \end{defn}
\noindent When $\s$ is boundless and sanitized by the regular Laplace mechanism, then $\s^*$ is unbiased for $\s$ since  $\E(\s^*)=\s$ per the definition of the Laplace distribution. If $\delta_1\propto n^{-k}$, where $k>0$, then   $\s^*$ is also consistent for  $\s$. When $\s$ is bounded and sanitized via the truncated or BIT Laplace mechanism, we will have biased $\s^*$ unless the bounds are symmetric around the original $\s$.  Proposition \ref{thm:mu12} states the formal conclusions and presents the magnitude of the bias of $s^*$  in the truncated and BIT Laplace mechanisms. The proof of Proposition \ref{thm:mu12} is provided in Appendix \ref{app:mu12}.
\begin{pro} 
[\textbf{bias of sanitized statistic from truncated  and BIT Laplace mechanisms}]\label{thm:mu12} Let $[c_0,c_1]$ be the global bounds on a singular $s$, $\lambda$ be the scale parameter, $s$ be the location parameter of the Laplace distribution, $\mu_1$ be the expected mean of the truncated Laplace distribution $f(s^*|s^*\!\in\![c_0, c_1])$, and $\mu_2$ be the expected mean of the BIT Laplace distribution defined in Eqn (\ref{eqn:BIT}). Then
\hspace{-2.5pt}
\begin{align}
\mu_1&=s+\frac{\frac{\lambda-c_0+s}{2}\exp\left(\frac{c_0-s}{\lambda}\right)-\frac{\lambda+c_1-s}{2}\exp\left(\frac{s-c_1}{\lambda}\right)}
{1-\frac{1}{2}\exp(\frac{c_0-s}{\lambda})-\frac{1}{2}\exp(\frac{s-c_1}{\lambda})}\label{eqn:mu1}\\
\mu_2&= s+\frac{\lambda}{2}\left[\exp\left(\frac{c_0-s}{\lambda}\!\right)-\exp\left(\frac{s-c_1}{\lambda}\right)\right] \label{eqn:mu2}
\end{align}
\textbf{(a)} $\mu_1\!=\!\mu_2\!=\!s$ ($s^*$ is unbiased for $s$) if and only if $c_0+c_1\!=\!2s$ ($c_0$ and $c_1$ are symmetric around $s$).\newline
\textbf{(b)}  $(\mu_1-s)(\mu_2-s)\ge0$ (the direction of the bias is the same between the two). Specifically,  $\mu_1-s\ge0 \mbox{ and } \mu_2-s\ge0$ if $s-c_0<c_1-s$; $\mu_1-s\le0  \mbox{ and } \mu_2-s\le0$ if $s-c_0>c_1-s$.\newline
\textbf{(c)} $|\mu_1-s|\ge|\mu_2-s|$  ($s^*$ sanitized via the BIT Laplace sanitizer is no more biased than that via the truncated Laplace  sanitizer).
\end{pro}

For  the global and data invariant bounds $[c_0,c_1]$, it is unlikely the sanitized results would be  unbiased via the truncated or the BIT Laplace mechanism per part (a) of Proposition \ref{thm:mu12}  as  $[c_0,c_1]$ are fixed while the original statistic $s$ changes from data to data. To achieve unbiasedness for $s^*$, local  bounds that depend on specific data sets can be constructed at additional privacy cost.   For example, bounds $[s-\min(s-c_0,c_1-s), s+\min(s-c_0,c_1-s)]$, which are symmetric around $s$,  can be used to bound sanitized results in the truncated and BIT Laplace mechanism that leads to unbiased $s^*$.  However, since the bounds are  functions of the original $s$, they will leak information about $s$, which has to be counted for in the total privacy cost.

\begin{rem}[\textbf{Extension of Proposition \ref{thm:mu12} to multidimensional $\s$}]\label{rem:multi1}
Proposition \ref{thm:mu12} examines a scalar query $s$. The conclusions can be easily extended to $\s_{r\time1}=(s_1,\ldots,s_r)$ of arbitrary dimension $r\ge1$ with univariale bounding constraint per statistic. The only difference lies in how the scale parameters $\lambda$ is defined. As given in Eqn (\ref{eqn:laplace}), the Laplace mechanism for sanitizing $\s_{r\times1}$ is $\prod_{j=1}^r\mbox{Lap}(s_j,\delta_1\epsilon^{-1})$  with $\delta_1=\sum_{j=1}^r\delta_{j1}$, where $\delta_{j1}$ is the global sensitivity of $s_j$, the $j$-th statistic in $\s$. We can rewrite The joint distribution of sanitized $\s^*$ as 
\begin{equation}
\prod_{j=1}^r\mbox{Lap}\left(s_j,\delta_{1j}\left(\epsilon\frac{\delta_{1j}}{\delta_1}\right)^{-1}\right) 
= \prod_{j=1}^r\mbox{Lap}(s_j,\delta_{1j}(w_j\epsilon)^{-1}),
\mbox{ where } w_j=\delta_{1j}/\delta
\end{equation}
In other words, the Laplace mechanism can be regarded as a special application of the sequential composition \citep{mcsherry2009privacy} by allocating a portion $w_j=\delta_{1j}/\delta$  of total $\epsilon$ to the sanitization of $s_j$ (also refer to \citet{liu2017}). For the scenario examined in this paper, that is, the bounds for $\s$ are publicly known and fixed, the conclusions from Proposition \ref{thm:mu12} apply to each element $s_j$ in $\s_{r\times 1}$ with different bounds $[c_{0j},c_{1j}]$ and scale parameter $\lambda_j=\delta_j(w_j\epsilon)^{-1}$ for different $s_j$ in Eqns (\ref{eqn:mu1}) and (\ref{eqn:mu2}).
\end{rem}

\begin{rem}[\textbf{Proposition \ref{thm:mu12} for one-sided bounding constraints}]
The conclusions in Proposition \ref{thm:mu12} are also applicable in the special cases when the bounds are one-sided $[c_0,\infty]$ or $[\infty,c_1]$, if such needs exist in practice. The proof and establishment of these conclusions do not impose any restrictions on the values $c_0$ and $c_1$ except for the trivial requirement $c_0<c_1$. The bias terms actually have  simpler expression in the case. Specifically,  Eqns (\ref{eqn:mu1}) and (\ref{eqn:mu2}) become
\begin{align}
\mbox{bounds } [c_0,\infty): &\; \mu_1=s+\frac{\frac{\lambda-c_0+s}{2}\exp\left(\frac{c_0-s}{\lambda}\right)}
{1-\frac{1}{2}\exp(\frac{c_0-s}{\lambda})};\; 
\mu_2= s+\frac{\lambda}{2}\exp\left(\frac{c_0-s}{\lambda}\!\right) \label{eqn:c0}\\
\mbox{bounds } (-\infty,c_1]: &\;
\mu_1=s-\frac{\frac{\lambda+c_1-s}{2}\exp\left(\frac{s-c_1}{\lambda}\right)}
{1-\frac{1}{2}\exp(\frac{s-c_1}{\lambda})};\;
\mu_2= s-\frac{\lambda}{2}\exp\left(\frac{s-c_1}{\lambda}\right).\label{eqn:c1}
\end{align}
It is easier (compared to the general case of $[c_0,c_1]$) to see that both biases from the truncation and the BIT procedures are positive when the bounds are $[c_0,\infty)$,  a special case of $s-c_0<c_1-s$ in part b), and negative  when the bounds are $(-\infty,c_1]$， a special case of $s-c_0>c_1-s$.  The difference of the two bias terms  when the bounds are $[c_0,\infty]$ is $\left[\frac{1-(c_0-s)/\lambda}
{1-\exp(\frac{c_0-s}{\lambda})/2}-1\right]\frac{\lambda}{2}\exp\left(\frac{c_0-s}{\lambda}\!\right)$. Since  $0<1-\exp(\frac{c_0-s}{\lambda})/2<1/2$ and  $1-(c_0-s)/\lambda>1$, thus the bias difference is $>0$. Since both biases are positive, this implies that $\mu_1$ has a larger positive bias than $\mu_2$. Similarly, when the bounds are $[-\infty, c_1]$, the bias difference between $\mu_1$ and $\mu_2$ is  $\left[1-\frac{1+(c_1-s)/\lambda}
{1-\exp(\frac{s-c_1}{\lambda})/2}\right]\frac{\lambda}{2}\exp\left(\frac{s-c_1}{\lambda}\!\right)$.  Since  $0<1-\exp(\frac{c_1-s}{\lambda})/2<1/2$, and  $1+(c_1-s)/\lambda>1$, the bias difference is $<0$. Since both biases are negative, this implies that $\mu_1$ has a larger negative bias than $\mu_2$. Taken together, $|\mu_1-s|>|\mu_2-s|$. 
\end{rem}

In addition to quantify the magnitude of bias, it is also of interest to bound the error of a sanitized statistic relative to its original value. Proposition \ref{lem:error} examines the upper bound for the mean squared error (MSE) for a sanitized statistic via the truncated and BIT Laplace mechanisms, and establishes the condition under which the statistic would be MSE-consistent for its original value, and the convergence rate of consistency.
\begin{pro}[\textbf{consistency and convergence rate}]\label{lem:error}  Let $s$ be the location parameter and $\lambda$ be the scale parameter of the Laplace distribution, and $s^*$ be the sanitized result for $s$ via  the truncated  or the BIT Laplace mechanism. The MSE $\E_{s^*}(s^*-s)^2$ is upper bounded by $2\lambda^2 = 2(\delta_1/\epsilon)^2$ and  $s^*$ is MSE-consistent for  $s$ as  $\lambda\rightarrow 0$. If $\delta_1\propto n^{-k}$, where $k>0$ and $n$ is the sample size, then the rate of the MSE converging  to 0  is $O(n^{-2k})$ for a given $\epsilon$.
\end{pro}
The proof is provided in Appendix \ref{app:error}.  The proof does not have restrictions on the values of $c_0$ or $c_1$, which can be as extreme as $c_1=\infty$ and $c_0=-\infty$. Therefore, the conclusions in Proposition \ref{lem:error} also to the non-sided bounding constraints. In addition, similar to Remark \ref{rem:multi1} for
Proposition \ref{thm:mu12}, the established conclusions in  Proposition \ref{lem:error} can be easily extended to $\s_{r\time1}=(s_1,\ldots,s_r)$ of arbitrary dimension $r\ge1$ with univariate bounding constraint. The only difference is that the statistic-specific $\lambda_j$ ($j=1,\ldots,r$  would be used. Therefore, different elements from $\s_{r\time1}$ will have different MSE bounds and different convergence rates. 

Proposition \ref{lem:error} shows that  the sanitized result, though likely biased per Proposition \ref{thm:mu12},  can still enjoy nice asymptotic properties such as asymptotic unbiasedness and consistency as $\lambda\rightarrow0$ . In the framework of the truncated  and BIT Laplace mechanisms, the scale parameter of the associated Laplace distribution $\lambda$ is $\delta_1\epsilon^{-1}$. For a pre-specified $\epsilon$, to satisfy the condition $\lambda\rightarrow0$, then $\delta_1$ needs to be $\rightarrow0$.  Intuitively speaking, as $n$ increases, the influence of a single individual on an aggregate measure of a data set is likely to diminish, and the individual is less prone to be identified from releasing the aggregate measure. Translated to the global sensitivity of the aggregate measure, it means that $\delta_1$ decreases with $n$.  $\delta_1$ of some commonly used statistics are $\propto n^{-1}$, such as proportions, means, variances and covariances (refer to the online supplemental materials), and  the  sanitized copies of these statistics via either the truncated or BIT Laplace mechanism are consistent for their original values by part c) of Proposition \ref{thm:mu12}. 
The results in Proposition \ref{lem:error} imply that the MSE of a sanitized statistic from the truncated or the BIT Laplace mechanism is comparable to the MSE from the regular Laplace mechanism without bounding, which is also $2\lambda^2$. In other words, the bounding does not seem to affect the MSE of a sanitized statistic despite the loss of unbiasedness in general. 

Propositions \ref{thm:mu12} and  \ref{lem:error} examine the statistical properties of the sanitized $\s^*$ relative to the original $\s$.  In many statistical analysis, the ultimate goal is to infer unknown population parameters $\theta$ based on the sample data.  Suppose $\s$ is an estimator for parameter $\boldsymbol{\theta}$ in a statistical model. Proposition \ref{prop:asymp} lists the sufficient conditions for $\s^*$, the sanitized version of $\s$, to be asymptotically unbiased and consistent for $\boldsymbol{\theta}$.
\begin{pro}[\textbf{asymptotic unbiasedness and consistency of sanitized statistic for population parameter}]\label{prop:asymp}
\textcolor{white}{.}\newline
(a) If $\E(\s|\boldsymbol{\theta})=\boldsymbol{\theta}$ or  $\E(\s|\boldsymbol{\theta})\rightarrow\boldsymbol{\theta}$, and if $\E_{\s^*}(\s^*|\s)\rightarrow \s$ as $n\rightarrow\infty$, then $\s^*$ is asymptotically unbiased for $\theta$; that is, $\E(\s^*|\boldsymbol{\theta})\rightarrow\boldsymbol{\theta}$.\newline
(b) If $\s^*\xrightarrow{p}{}\s$ and $\s\xrightarrow{p}{}\boldsymbol{\theta}$ as $n\rightarrow\infty$, then $\s^*$ is consistent for $\theta$; that is, $\s^*\xrightarrow{p}{}\boldsymbol{\theta}$.
\end{pro}
\noindent  The proof of Proposition \ref{prop:asymp} is given in Appendix \ref{app:lem}. Note that Proposition \ref{prop:asymp} does not list the conditions for obtaining an unbiasedness $\s^*$ for $\boldsymbol{\theta}$ as it is meaningless given the low  likelihood of obtaining an unbiased $\s^*$ for $\s$ per Proposition \ref{thm:mu12}.  Proposition \ref{prop:asymp} implies that asymptotic unbiasedness and consistency of the sanitized  $\s^*$ for $\boldsymbol{\theta}$ can be achieved in two steps. In step one, we will choose an estimator $\s$ that is asymptotically  unbiased or consistent for $\boldsymbol{\theta}$, which should be relatively straightforward given that these types of estimators are well studied and widely applied in statistics; in the second step, we will employ   an appropriate differentially private mechanism  to  generate $\s^*$ that is asymptotically  unbiased  or consistent for $\s$, such as the BIT and truncated Laplace mechanisms, which yield MSE-consistent $\s^*$  for $\s$ under the conditions listed in Propositions \ref{lem:error}.

\section{Simulation Studies} \label{sec:simulation}
We conducted two simulation studies to demonstrate the applications of the truncated and BIT bounding mechanisms and examine the statistical properties of the sanitized results.  In the first simulation, we sanitized a variance-covariance matrix, and focused on the comparisons between the sanitized results and the original results and  between the truncated and BIT truncated Laplace mechanisms on the effects on the sanitized results. In the second simulation, we sanitized proportions and focused on the inferential properties of the sanitized proportions by examining the bias, root mean squared error (RMSE) and coverage probability (CP) for the true proportions based on the sanitized results.

\subsection{Simulation Study 1}
In this simulation, we applied the truncated and BIT Laplace mechanisms to sanitize a  variance-covariance matrix $\mathbf{S}$ in a data set of size $n$. The variance-covariance matrix is a widely used statistic for examining the dependency structure among  multiple continuous variables.  It is also an ideal statistic to examine the effects of the two mechanisms given that every element in the matrix has to satisfy some type of bounding constraints. The bounding constraints of a covariance matrix of any dimension include that the marginal variances are positive and the correlations are bounded between [-1, 1]. Additionally,  the marginal variances are  right-bounded in bounded data from which $\mathbf{S}$ is calculated. Table \ref{tab:sim2} summarizes the bounds and global sensitivity of the components in $\mathbf{S}$.
\begin{table}[!htb]
\caption{Global sensitivity of variance and covariance terms in a covariance matrix}\label{tab:sim2}
\centering\begin{tabular}{rlll}
\hline
statistic & &  bounds$^\ddag$ & $\delta_1$\\
\hline
variance & $S_{jj}$  & $(0, n(c_{j1}-c_{j0})^2/(4(n-1))]$      & $(c_{j1}-c_{j0})^2/n$  \\
variance & $S_{j'j'}$   & $(0, n(c_{j'1}-c_{j'0})^2/(4(n-1))]$      & $(c_{j'1}-c_{j'0})^2/n$  \\
covariance & $S_{jj'}$ & $(-\sqrt{S_{jj} S_{j'j'}},\sqrt{S_{jj} S_{j'j'}})$     & $(c_{j1}-c_{j0})(c_{j'1}-c_{j'0})/n$  \\
\hline
\multicolumn{4}{l}{\footnotesize{$^\ddag [c_{j0},c_{j1}]\times[c_{j'0},c_{j'1}]$ were the bounds of variables $X_j$ and $X_{j'}$}}\\
\end{tabular}
\end{table}

When sanitizing $\mathbf{S}$ in general, we first obtain legitimate sanitized $S_{jj}^*$ and $S_{j'j'}^*$, and then sanitize $S_{jj'}$ given  $S_{jj}^*$ and $S_{j'j'}^*$ under the constraint that $S^*_{jj'}\in [-\sqrt{S^*_{jj} S^*_{j'j'}},\sqrt{S^*_{jj} S^*_{j'j'}}]$.  Though the bounds for $S^*_{jj'}$ depend on $S^*_{jj}$ and $S^*_{j'j'}$,  the latter two are already sanitized; therefore, bounding procedures for $S^*_{jj'}$ using information $S^*_{jj}$ and $S^*_{j'j'}$    does not incur additional privacy cost.  The interval constraint on $S^*_{jj'}$ does not have to contain the original $S_{jj'}$. Comparing to containing $S_{jj'}$, which would be nice from information preservation perspective, satisfying the hardcore requirement that the Pearson correlation is within $[-1, 1]$ is necessary, when there is a conflict between the two. In addition, requiring the interval to contain the original $S_{jj'}$ would also lead to privacy leak, strictly speaking.

It is possible that the sanitized covariance matrix $\mathbf{S}^*$ is not positive definite (PD) with the element-wise sanitization approach.  If a sanitized covariance matrix is not PD and has a significant number of (small) negative eigenvalues, it can be made PD with semi-definite optimization via, e.g.,  the alternating projections algorithm \citep{Higham02}, the Newton methods for nearest correlation matrix \citep{Qi,Higham10a}, and the spectral projected gradient method \citep{Qi,Higham10b}\footnote{The R function \texttt{nearPD()}  in package \texttt{Matrix} implements the alternating projections algorithm.}. A possible alternative is to sanitize $\mathbf{S}$ as a whole instead of element-wise, an interesting  and worthwhile topic for future research.

We examined in this simulation study a $2\times2$ variance-covariance matrix with three different specifications of $(S_{11},S_{22},r)$: $(1,2,0), (1,2,-0.4)$, and $(1,2,0.7)$, respectively.  The 3 correlation settings allow us to examine the bounding effects on the pairwise correlation when there is no correlation, moderate (negative) correlation, and strong (positive) correlation. We set the global bounds  $[c_{10},c_{11}]$ at $[-3,3]$ and $[c_{20},c_{21}]$  at  $[-4.5,4.5]$ \footnote{For approximately Gaussian variables with means 0, both the bounds of $[-3,3]$ with a standard deviation of 1  and $[-4.5, 4.5]$ with a standard deviation of $\sqrt{2}$  represent $>99\%$ data mass though this simulation did not require the Gaussian assumption.}. The total privacy budget was  $\epsilon=1$.  Since the 3 statistics were calculated on the same set of data, the  sequential composition principle applied  \citep{mcsherry2009privacy} when it comes to privacy budgeting. There are different ways to allocate the total budget when sanitizing multiple statistics calculated from the same data set,  such as by an equal allocation across all the statistics or according to some type statistical or practical ``importance'' of the statistics (see \citet{liu2017} for more discussion). Here we divided the total privacy budget $\epsilon$ equally among the 3 statistics; that is, each sanitization received $1/3$ of the total budget. We also investigated a wide range of sample size  $n$  from 50 to 800.  In each examined scenarios  of  $(\mathbf{S},n)$, 500 independent sanitizations were carried out to examine the distributional properties of the sanitized results.  

\begin{landscape}
\begin{figure}[!htb]
\centering \includegraphics[scale=0.7]{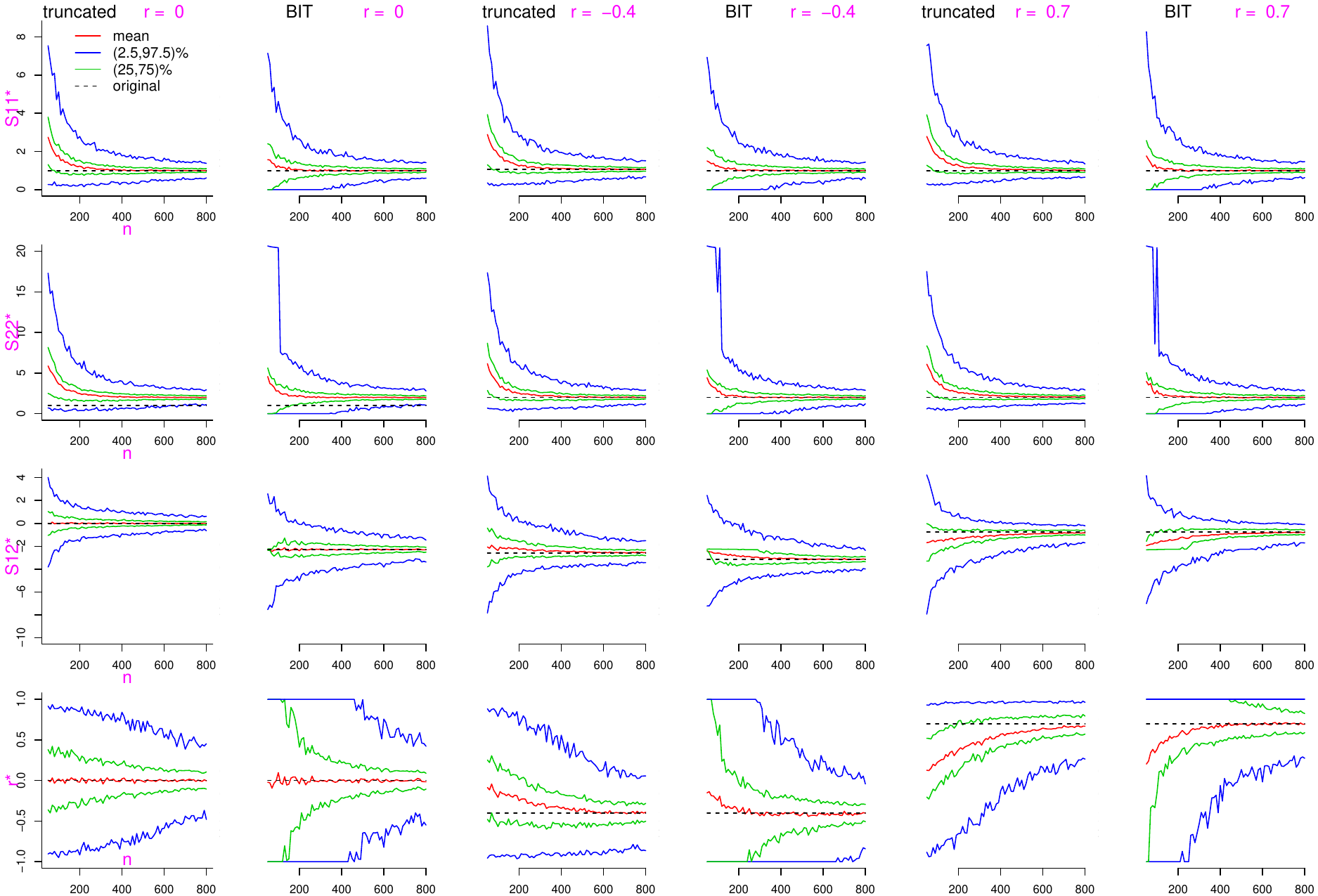}
\caption{Sanitized components in $\mathbf{S}$  ($(Q_1,Q_2)$\% represents the $Q_1$ and $Q_2$ percentiles of the distribution of the sanitized statistic)}\label{fig:sim2}
\end{figure}
\end{landscape}

The results are presented in Figure \ref{fig:sim2}. In each plot, the average and the (2.5\%, 25\%, 75\%  and 97.5\%) percentiles of the sanitized results from the 500 sanitizations are presented, benchmarked against the original results. The main findings are summarized as follows. First, when $n$ was relatively small, there was noticeable  bias in the sanitized results compared to the original results, except for $S_{12}$ and $r$ when $r=0$ (the boundaries were symmetric about the original results and thus there was no bias per part (a) of Proposition \ref{thm:mu12}).  Second, the sanitized results generated via the truncated Laplace mechanism were more biased  than those via the BIT Laplace mechanism, consistent with part (b) of Proposition \ref{thm:mu12}. Third,  as $n$ increased, both the bias and dispersion of the sanitized results approached 0, consistent with part (c) of Proposition \ref{thm:mu12}. Lastly,  when $n$ was small, the scale parameter associated with the Laplace distribution in both mechanisms was large, therefore more sanitized values were set at the boundary values in the BIT mechanism  and the distribution of the sanitized values became flatter in the truncated mechanism,  especially for  $r$.

\subsection{Simulation Study 2}
In this simulation, we aimed to release a proportion vector $\mathbf{p}$ where $\sum_{j=1}^K p_j=1$. Proportions are very common statistics in public data release.  For example, $\mathbf{p}$ could be the proportions of different income levels in the US population, or the cell proportions from  a cross-tabulation of, say gender and race. Besides the bounding constraint $[0,1]$ on each proportion component, $\mathbf{p}$ is also subject to the equality constraint $\sum_{j=1}p_j=1$, which has to be retained in the released sanitized results, making it an interesting problem to study.

Since the proportions are calculated from disjoint subsets of data, the addition or removal of a single observation affects the count in exactly one cell,  the global sensitivity ($\delta_1$) of releasing $\mathbf{p}$ is $n^{-1}$ or $2n^{-1}$, depending on which definition of ``differing by one record'' is used on $\Delta(x,x')=1$ (refer to the online supplementary materials).  With the Laplace mechanism, sanitization of each proportion in  $\mathbf{p}$ is perturbed with a noise term from Lap($\delta_1\epsilon^{-1}$). In this simulation, we used $\delta_1=n^{-1}$,  and examined 3 different specifications of $\epsilon$ ($0.1, 0.5, 1$) and a range of sample size ($n=50 \sim 500$) (the results obtained  from this simulation are also applicable to $\delta_1=2n^{-1}$  with doubled $\epsilon$). 500 multinomial data sets, each sized at $n$, were simulated from multinomial$(n, \mathbf{p})$. We examined a 4-element $\mathbf{p}$ in this simulation and set $\mathbf{p} =(0.1,0.2,0.3,0.4)$ (these parameter values were chosen because they expand a wide range -- some are closer to the boundaries while others are closer to the center, allowing us to examine the effects of the binding mechanisms on proportions of different magnitude).

The sample proportions $\hat{\mathbf{p}}$ were calculated in each repeat and were sanitized via the truncated and BIT Laplace mechanisms respectively. We employed 3 procedures to ensure the equality constraint  $\sum_{k=1}^4 p_{k}=1$ was met (in addition to the bounding constrained on each proportion). In the first approach (re-scaling and normalization),  each proportion in $\hat{p}_j$ was sanitized independently to obtain the perturb version $\hat{q}^*_{k}$, which was then  normalized to obtain $\hat{p}^*_j=\hat{q}^*_{k}(\sum_{k=1}^4 \hat{q}^*_{k})^{-1}$;  it is $\mathbf{\hat{p}}^*$ that was released. This re-scaling approach was intuitive and straightforward.  In the second approach, referred to as the all-but-one approach,  we sanitized 3 proportions out of 4, and then calculated the 4-th proportion via $1-\sum_{k=1}^3 \hat{q}^*_{k}$.  The all-but-one approach obeyed the equality constraint during the sanitization without post-processing as in the re-scaling approach.  In the third approach, referred to as  the universal histogram/UH approach, we applied a slightly modified  procedure  from \citet{boosting} to ensure the equality constraint in $\mathbf{p}$.  In the original UH procedure in \citet{boosting}, the root node is sanitized and there is no inequality constraint; in our case, the root node was  fixed at 1 as it was the summation of $\mathbf{p}$, and there were inequality constraints. Since the UH requires the number of children per node to be constant across the whole tree and there needs to be at least three layers in the tree  (since the root node is fixed at one in our case) in order to show any level of improvement in the accuracy of some of the sanitized nodes, the four proportions in $\mathbf{p}$, represented by four leaf nodes, are thus combined in a binary fashion in 3 layers; otherwise, it would be the same as the re-scaling approach.
\begin{enumerate}[leftmargin=*,align=left] \setlength\itemsep{0em}
\item Arrange the 4 proportions in a 3-layer binary tree structure. The root node in the tree always has a value of 1, and its two child nodes $h_1$ and $h_2$ in layer 2 satisfy the constraint $h_1+h_2=1$ (equality constraint 1). Similarly, the two child nodes of $h_1$ ($h_{11}$ and $h_{12}$ in layer 3) satisfy $h_{11}+h_{12}=h_1$  (equality constraint 2) and the two child nodes of $h_2$ ($h_{21}$ and $h_{22}$ in layer 3) satisfy $h_{21}+h_{22}=h_2$  (equality constraint 3). As a result, the four leaf nodes in layer 3, corresponding to the 4 proportions in $\mathbf{p}$, satisfy $h_{11}+h_{12}+h_{21}+h_{22}=1$.
\item Sanitize $h_1$ and $h_2$ in layer 2, and $(h_{11},h_{12},h_{21}, h_{22})$ in layer 3 via the BIT and Laplace Laplace mechanisms with the scale  parameter $2/(n\epsilon)$. The global sensitivity doubles in this procedure as there are two sets of proportions to be sanitized on the same data set.
\item Calculate the inconsistent counts $\mathbf{z}$ (due to  sanitization) for the nodes in the tree. Let $\mathbf{h}^*(v)\!=\!\{h^*_1,h^*_2,h^*_{11},h^*_{12},h^*_{21},h^*_{22}\}$ denote the sanitized results from step 2. $z(v)=h^*(v)$ if $v$ is a leaf node in layer 3, and $z(v)=\frac{1}{3}\sum_{v'\in\mbox{\footnotesize{children of }}v}z(v')+ \frac{2}{3}h^*(v)$ for $v$ in layer 2.
\item Correct the inconsistency in the tree. For nodes in layers 2 and 3,  the corrected nodes $\bar{h}(v)\!=\! z(v)\!+\!\frac{1}{2}(h(u)\!-\!\!\sum_{v'}\!z(v'))$, where $u$ is the parent of $v$, and $\bar{h}(v)\!=\!1$ for the root node\footnote{Even if the BIT or the truncated Laplace mechanisms is employed in Step 2, negative nodes might re-appear after the correction in Step 4. If that occurs, the negative nodes are set at 0 and the nodes that share the same parent node with the negative nodes are re-normalized to sum up to their parent node. We also tried applying the regular Laplace mechanism directly in Step 2 and used the  BIT or truncated Laplace mechanisms to bound the nodes after Step 4, and the results were similar or slightly worse compared to the above procedure, depending on which statistics to release.}. 
\item Release $\bar{\mathbf{h}}(v)$.
\end{enumerate}
\citet{boosting}  stated that the UH procedure optimizes the accuracy of the sanitized nodes closer to the root (low-order marginals)  with the smallest MSE (relative to the original results) among the approaches that yield unbiased sanitized estimators for the original results while satisfying the equality/consistency constraint. However, the UH procedure decreases the accuracy of the sanitized high-order nodes and the leaf nodes. This implies the accuracy of the 4  individual proportions in $\mathbf{p}$, which were the leaf nodes in the simulation,  will suffer, but some linear combinations of $\mathbf{p}$, say $p_1+p_2$ might have higher accuracy than from the re-scaling and the all-but-one approaches.  In addition, the UH procedure can be sensitive to the order of how the tree is built. We tried two different ways of grouping the 4 proportions into two nodes ($h_1$ and $h_2$) in the second layer. One way  ($(h_{11}=p_1=0.1,h_{12}=p_2=0.2)\in h_1;(h_{21}=p_3=0.3,h_{22}=p_4=0.4)\in h_2)$ seemed slightly better in preserving the original information than the other ($(h_{11}=p_1,h_{12}=p_4)\in h_1;(h_{21}=p_2,h_{22}=p_3)\in h_2)$; therefore, we only present the results from the former.

We calculated the bias and RMSE relative to the true $\mathbf{p}$, and the coverage probability (CP) of the 95\% confidence interval for the true $\mathbf{p}$ based on the sanitized $\hat{\mathbf{p}}^*$ in each of the 3 approaches. The bias, RMSE, and CP based on the sanitized results were compared to those based on the original $\mathbf{p}$. As discussed above, some of the lower-order nodes (such as $h_1$ and $h_2$) might have improved accuracy in the UH approach, we thus also compared the inferences of $p_1+p_2$, $p_2+p_3$ and $p_2+p_4$ between the re-scaling and the UH approach. There is no need to examine the sum of 3 proportions, say $p_1+p_2+p_3$, as its accuracy would be the same as the individual proportion $p_4$ under the equality constraint. 

In the re-scaling approach, there was minimal bias in sanitized $\hat{\mathbf{p}}^*$ when $\epsilon=1$ and $\epsilon=0.5$ across all $n$ and both bounding mechanisms (truncation or BIT); and there was some bias at small $n$ when $\epsilon=0.1$, especially for the smallest proportion $p_1=0.1$ (positive bias) and the largest proportion $p_4=0.4$ (negative bias).  Consistent with Proposition \ref{thm:mu12}, the BIT mechanism yielded less bias than the truncated mechanism. The RMSE of  the sanitized results was inflated compared to  the original RMSE, which was expected given the noise injected during the sanitization step. The larger $\epsilon$ or the larger $n$ was, the smaller the inflation was.  When $\epsilon=1$, the sanitized and the original RMSE values were basically the same. Though the BIT mechanism led to smaller bias compared to the truncated mechanism for  small $n$ when $\epsilon=0.1$,  the RMSE values were larger in the former. Finally,  the CP was  around 95\% at all $n$ when $\epsilon=1$, decreased to $85\%\sim92\%$ for $n\in[50,300]$ when $\epsilon=0.5$, and down to $50\%\sim80\%$ for all $n\in[50,500]$  when $\epsilon=0.1$. The BIT  mechanism had worse under-coverage than the truncated mechanism at small $n$  for $\epsilon=0.1$ and yielded similar CP as the truncated mechanism in other cases. 

The under-coverage observed can be resolved to some degree by using the multiple synthesis (MS) technique in DP \citep{liu2017, bowen}. The MS takes into the variability introduced by the sanitization process by releasing multiple synthetic sets. In this case, we independently sanitized  5 sets in each simulation scenario; and the inferences were then combined over the 5 sets using the rule given in \citet{liu2017}. In order to maintain $\epsilon$-DP overall, each set was sanitized using $1/5$ of the total privacy budget $\epsilon$ per the sequential composition theorem.  The results are given in Figure \ref{fig:sim1m}. The CP improved  significantly from releasing multiple synthetic data sets, especially for the BIT mechanism. However, due to the decreased privacy budget per synthetic set and the bounding, the sanitized results were  noisier and the biases were noticeably  larger, even after being averaged the 5 sets. For example,  there was noticeable bias at  $\epsilon=0.1$  and the RMSE did not approach the original RMSE within the examined range of $n$ for  both  mechanisms.

Compared to the re-scaling approach, 1) the all-but-one approach (Fig \ref{fig:sim1K-1}) had similar performance  in bias and RMSE for $\epsilon=0.5$ and 1; 2) when $\epsilon=0.1$, the bias was smaller at small $n$  but the RMSE was noticeably large in the all-but-one approach;3) the performance on CP was similar in all proportions but was much worse for $p_1=0.1$, which was the proportion calculated from the other 3.  
Compared to the re-scaling approach, the UH approach (Fig \ref{fig:sim1UH}) was worse in bias, RMSE and CP for all proportions at all $\epsilon$ values when $n$ was relatively small; the under-performance was the most obvious when $\epsilon=0.1$. The inferiority of the UH relative to the re-scaling approach are expected as the UH  benefits the accuracy of the low-order marginals (the nodes closer the root) but the leaf nodes (the individual proportions in $\mathbf{p}$) actually suffer a loss of accuracy.  

Figures S1 and S2 in the  supplementary materials  display how much improvement the UH approach brought to the pairwise sums in $\mathbf{p}$ compared to the re-scaling approach. The sanitized inferences on $p_1+p_2$ (node $h_1$) were better than  $p_2+p_3$ and $p_2+p_4$ (sum of two leaf nodes from different parents) in the UH approach. However, compared to the re-scaling approach, the accuracy of sanitized inferences on $p_1+p_2$ did not appear to be better; and those on   $p_2+p_3$ and $p_2+p_4$ were worse.  In summary, the UH approach was not efficient as the re-scaling approach in preserving the original information in the individual proportions in $\mathbf{p}$, and the supposed improvement in the low-level nodes  ($p_1+p_2$) was not obvious in this case either, perhaps due to 1) the additional inequality constraint, which is not considered in \citet{boosting}, on top of the equality constraint; and 2) the number of layers and nodes was not large enough to demonstrate the advantages of the UH approach.
 \begin{landscape}
\begin{figure}[H]
\centering \includegraphics[scale=0.64]{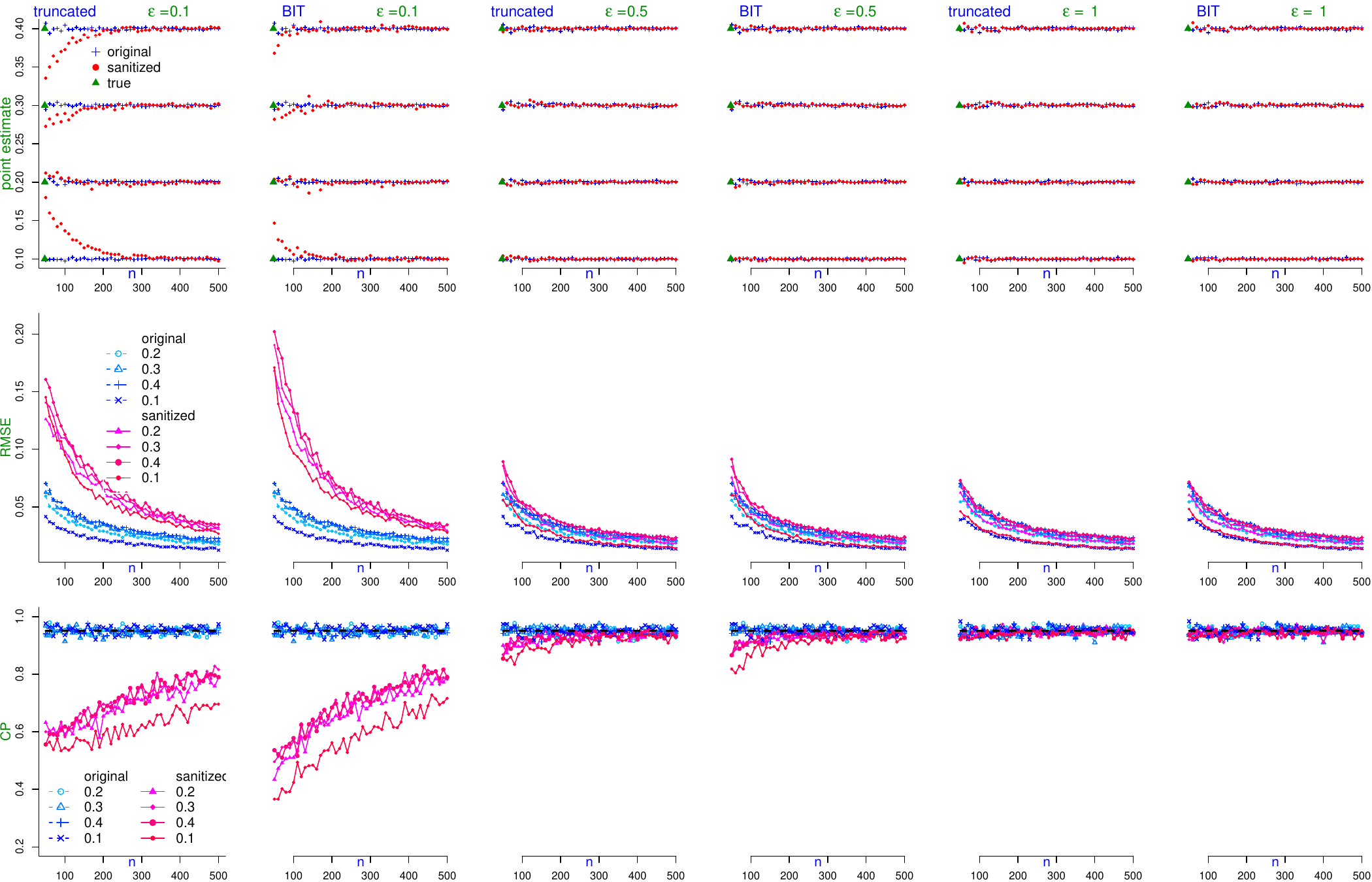}
\caption{Bias, RMSE and CP of sanitized proportions (the re-scaling approach was applied to satisfy the equality constraint; red lines represent the 4 original proportions, and blue lines represent the 4 sanitized proportions)}\label{fig:sim1}
\end{figure}
\begin{figure}[H]
\centering \includegraphics[scale=0.64]{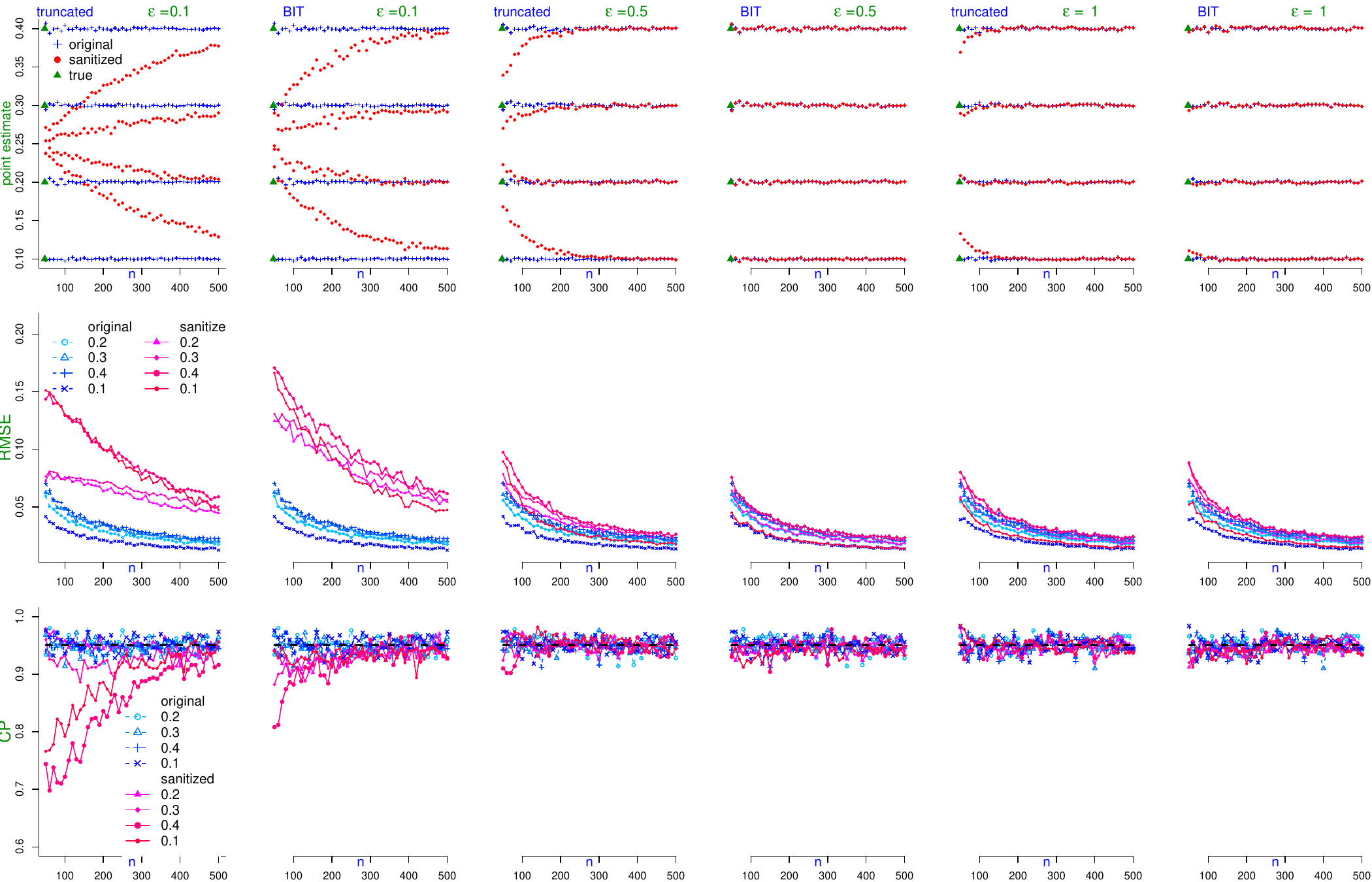}
\caption{Bias, RMSE and CP of sanitized proportions from multiple synthesis (the re-scaling approach was applied to satisfy the equality constraint; red lines represent the 4 original proportions, and blue lines represent the 4 sanitized proportions)}\label{fig:sim1m}
\end{figure}
\begin{figure}[H]
\centering \includegraphics[scale=0.64]{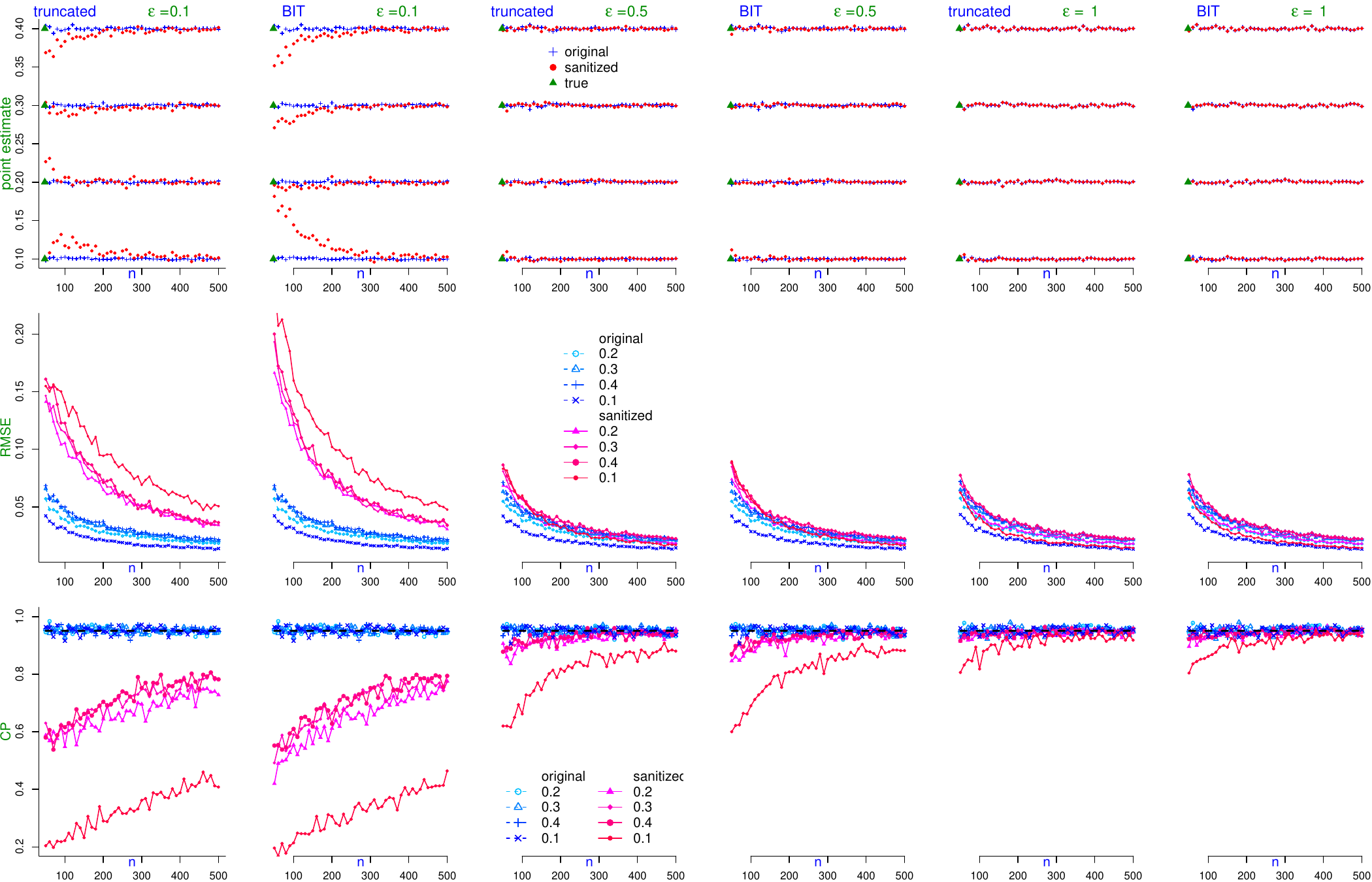}
\caption{Bias, RMSE and CP of sanitized proportions in the all-but-one approach (red lines represent the 4 original proportions, and blue lines represent the 4 sanitized proportions)}\label{fig:sim1K-1}
\end{figure}
\begin{figure}[H]
\centering \includegraphics[scale=0.64]{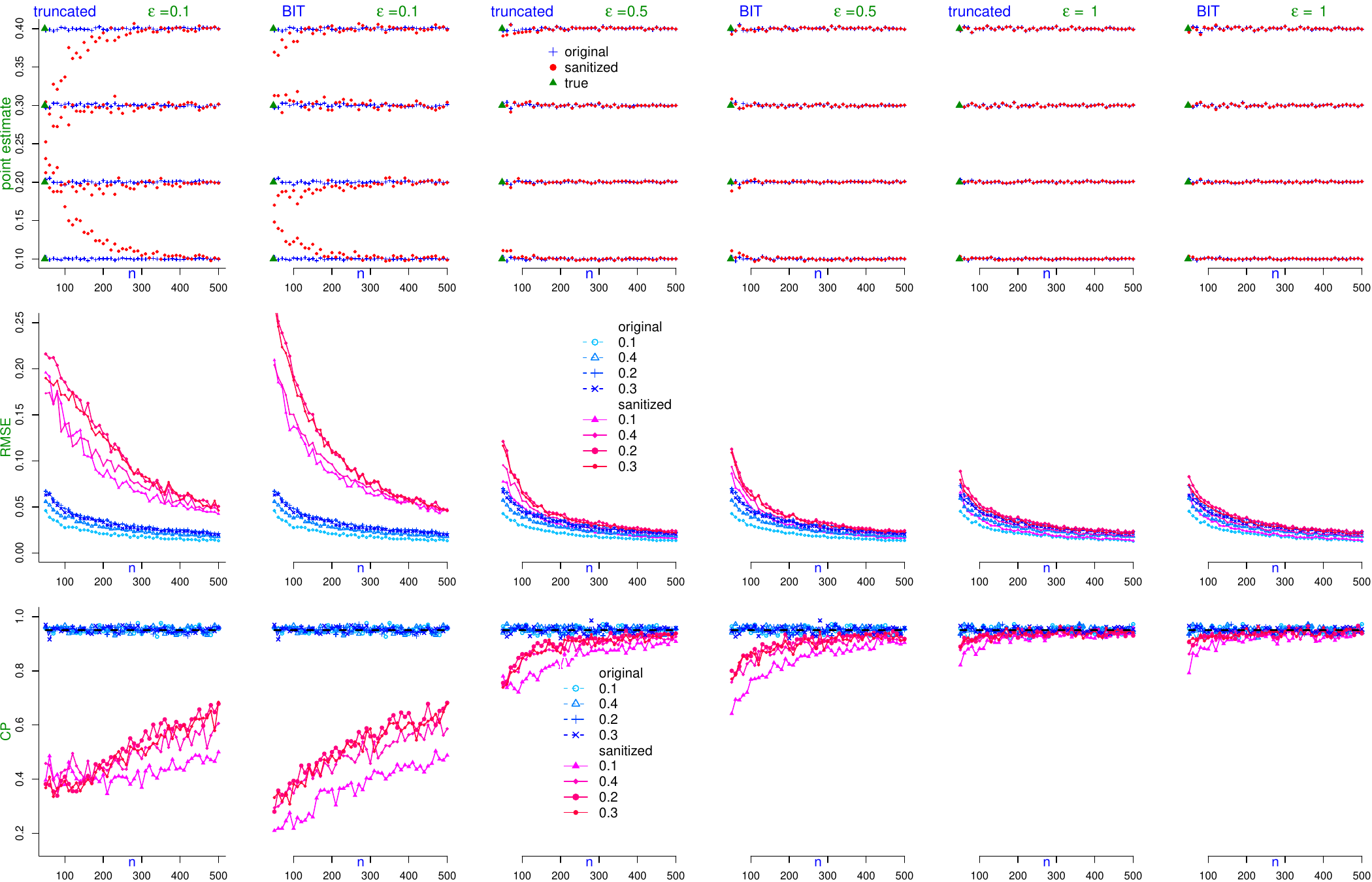}
\caption{Bias, RMSE and CP of sanitized proportions in a modified universal histogram procedure based on Hay et. al. (2010) (red lines represent the 4 original proportions, and blue lines represent the 4 sanitized proportions)}\label{fig:sim1UH}
\end{figure}
 \end{landscape}
 

\section{Discussion} \label{sec:discussion}
We defined two differentially private procedures for sanitizing statistics with bounding constraints and examined the statistical properties of  sanitized results from the two procedures. Both the  truncated and BIT Laplace procedures produce biased sanitized results relative to their original observed values unless the  bounds  are symmetric around the original results, which is a hard-to-satisfy condition in real life  given than the original statistics change by data  while the bounds are global and fixed. However, sanitized results can still be MSE-consistent for the original values if the scale parameter of the Laplace distribution associated with the two procedures approaches 0 as the sample size $n$ approaches $\infty$. We also provided an upper bound for the MSE between the sanitized results  relative the original for a definite $n$ in the truncated and BIT Laplace procedures.

Though the BIT Laplace mechanism in theory delivers less biased sanitized statistics than the truncated Laplace mechanism, the former does not seem to be more advantageous over the latter in practical applications for the following reasons. First, asymptotic unbiasedness and consistency hold in both procedures, so there is minimal difference between the two when $n$ is large. Second,  the truncated Laplace distribution  is a smooth distribution while the BIT Laplace distribution is piece-wise. Though the distributional shape might be irrelevant in the release of a single sanitized statistic,  it will when it comes to uncertainty quantification or making inferences about population parameters based on the sanitized data.  Last, the 3-piece distributional shape of the BIT Laplace distribution  requires the intervals of the outcomes to be closed on both ends so that the boundary values are exclusively defined. This is not necessary for the truncated Laplace distribution where the density function is continuous and smooth. This last point seems trivial but can be irritating in practical applications. For example, in the first simulation, closed intervals $[0, (c_1-c_0)^2 n/(4(n-1))]$ and $[-1,1]$ were applied to variance and correlation, respectively. As a result, some sanitized outputs were exactly 0 for variance, and  exactly -1 or 1 for  correlation from the BIT Laplace mechanism. In practice,  these values are rare occurrences due to measurement errors and noises, and the user may choose to reject the sanitized results that are valued exactly at the boundary. If the user demands more plausible results that agree  with  real-life situations, what values to replace the implausible boundary values becomes an arbitrary decision and could potentially affect the statistical properties of the sanitized results. Those concerns do not exist in the truncated Laplace mechanism.

This paper has focused on the truncation and BIT bounding procedures in the framework of the Laplace mechanism of $\epsilon$-DP. The  truncated and BIT  procedures are general enough to be extended to other differentially private sanitizers that output numerical results. For example, the Gaussian mechanism of $(\epsilon,\delta)$-aDP \citep{dwork2013algorithmic} or of $(\epsilon,\delta)$-pDP \citep{liu2016a} outputs sanitized values spanning the range $(-\infty,\infty)$, which be post-processed, if there are  bounding constraints  by the truncation or the BIT procedures to satisfy the constraints.   The statistical properties of sanitized results in other DP mechanisms other than the  the Laplace mechanism  will need to examined on  a case-by-case basis as the probability distribution of the injected noises are different.

Our work is different from \citet{qDP} though somewhat related. \citet{qDP} also propose the concept of the unbiasedness for the true query results after the directly sanitized results are post-processed (or “refinement” as they refer to) based on known background knowledge and constraints. However, their theoretical work focuses on linear constraints and some of these constraints contain the original non-private information. Therefore, they examine ``how refinement strategies affect differential privacy''. In other words, they study DP given sanitized results that satisfy non-private linear constraints. This is very different from our work, where the bounding and inequality constraints are public knowledge and do not contain any sensitive or  private information. Therefore, DP is preserved after post-processing and our investigation focus is how satisfying the public constraints affects the utility of the sanitized  results.
 
As briefly mentioned in the Introduction section,  releasing differentially private statistics with the bounding constrains falls can be dealt with the constrained inferences approach by minimizing the loss function measuring the deviance between constrained and unconstrained sanitized results while satisfying a defined set of constraints.  The simulation study implemented 3 approaches for satisfying the equality constraints, with one of them developed in the framework of the constrained inference, among a set of proportions that were also subject to bounding constraints. The simulation exercise only represents a small empirical attempt in handling constraints when releasing sanitized statistics; further research is warranted on the development of innovative and efficient approaches that release the optimal sanitized results under both inequality and equality constraints. Finally, we briefly touched on the topic of multivariable constraints throughout the discussion, including a simulation study on the multivariable linear equality constraint. More in-depth theoretical and empirical investigation on how multivariable constraints affect the statistical properties of sanitized results will be conducted in the future.


\section*{\normalsize{Supplementary Materials}}
The online supplementary materials contain the  calculations of the $l_1$ global sensitivity of some common statistics, including the sample proportion, mean, variance, and covariance; as well as  additional results from Simulation Study 2. The materials are available at \url{https://www3.nd.edu/~fliu2/bounding-suppl.pdf} \label{sec:sm}.

\section*{\normalsize{Acknowledgement}}
We thank two anonymous referees and the editor for their careful reviews and insightful and constructive comments, which helped improve the quality of the manuscript.

\section*{Appendix}
\appendix
\titleformat{\chapter}[display]
  {\normalfont\large\bfseries}
  {\chaptertitlename\ \thechapter}
  {20pt}
  {\large}
\numberwithin{equation}{section}

\section{\large{Proof of Proposition \ref{thm:mu12}}}\label{app:mu12}
The mean of a truncated Laplace distribution Lap$(s,\lambda|x\in[c_0,c_1]$ is 
\begin{align*}\textstyle
\mu_1=&\E(x|x\in(c_0,c_1))
=(F(c_1)-F(c_0))^{-1}\int_{c_0}^{c_1} \frac{x}{2\lambda}\exp\left(-\frac{|x-s|}{\lambda}\right)dx\\
\textstyle=&(F(c_1)-F(c_0))^{-1}\left[
\int_{c_0}^s \frac{x}{2\lambda}\exp\left(\frac{x-s}{\lambda}\right)dx + \int_s^{c_1} \frac{x}{2\lambda}\exp\left(\frac{s-x}{\lambda}\right)dx\right]\\
=&(F(c_1)-F(c_0))^{-1}\left(\textstyle s+\frac{1}{2}\!\left(\!(\lambda-c_0)\exp\!\left(\!\frac{c_0-s}{\lambda}\!\right)\!-\!(c_1+\lambda)\exp\!\left(\!\frac{s-c_1}{\lambda}\!\right)\!\right)\right)\\
=&(1-\frac{1}{2}\exp(\frac{s-c_1}{\lambda})- \frac{1}{2}\exp(\frac{c_0-s}{\lambda}))^{-1}\left(\textstyle s+\frac{1}{2}\!\left(\!(\lambda-c_0)\exp\!\left(\!\frac{c_0-s}{\lambda}\!\right)\!-\!(c_1+\lambda)\exp\!\left(\!\frac{s-c_1}{\lambda}\!\right)\!\right)\right)\\
=&s+\textstyle\left(\!1-\frac{1}{2}\exp(\frac{s-c_1}{\lambda})- \frac{1}{2}\exp(\frac{c_0-s}{\lambda})\!\right)^{-1}\!\!
\left(\frac{1}{2}(\lambda-c_0+s)\exp\!\left(\!\frac{c_0-s}{\lambda}\!\right)\!-
\frac{1}{2}(c_1+\lambda-s)\exp\!\left(\!\frac{s-c_1}{\lambda}\!\right)\!\right). \newline
\end{align*}
The mean of the BIT Laplace distribution is $$\mu_2=p_0c_0+p_1c_1+(1-p_0-p_1)\E(x|x\in[c_0,c_1]).$$ 
Since $p_0=\Pr(x< c_0) =  \frac{1}{2}\exp(\frac{c_0-s}{\lambda})$ and  $p_1=\Pr(x> c_1) = \frac{1}{2}\exp(\frac{s-c_1}{\lambda})$, and given the result from Part a), then $\mu_2$ is
$\textstyle\frac{c_0}{2}\exp\left(\frac{c_0-s}{\lambda}\right)+\frac{c_1}{2}\exp\left(\frac{s-c_1}{\lambda}\right)+
s+\frac{\lambda-c_0}{2}\exp\!\left(\!\frac{c_0-s}{\lambda}\!\right)\!-\!\frac{c_1+\lambda}{2}\exp\!\left(\!\frac{s-c_1}{\lambda}\!\right)=
\textstyle s+\frac{\lambda}{2}\left[\exp\!\left(\!\frac{c_0-s}{\lambda}\!\right)-\exp\!\left(\!\frac{s-c_1}{\lambda}\!\right)\right].$

\noindent \textbf{Part a)}: In the case of $\mu_1$, $s^*$ is unbiased for $s$ if $(\lambda-c_0+s)\exp\!\left(\!\frac{c_0-s}{\lambda}\!\right)=
(c_1+\lambda-s)\exp\!\left(\!\frac{s-c_1}{\lambda}\right)$.    Let $f(x)= (\lambda +|x|)\exp(\!-\frac{|x|}{\lambda})$, where $x$ is a real number.  $f(x)$ is symmetric about $x=0$. $f'(x)=-\frac{|x|}{\lambda}\exp(\!-\frac{|x|}{\lambda})$; therefore, $f(x)$ is a monotonic increasing function when $x<0$ and a  monotonic decreasing function when $x>0$. Taken together, $(\lambda-c_0+s)\exp\!\left(\!\frac{c_0-s}{\lambda}\!\right)=
(c_1+\lambda-s)\exp\!\left(\!\frac{s-c_1}{\lambda}\right)$ and $s^*$ is unbiased for $s$  iff $c_0$ and $c_1$ are symmetric about $s$. In the case of $\mu_2$, $s^*$ is unbiased for $s$ if $\exp\!\left(\!\frac{c_0-s}{\lambda}\!\right)=\exp\!\left(\!\frac{s-c_1}{\lambda}\right)$.  $f(x)$ is symmetric about $x=0$ Let $f(x)=\exp(\!-\frac{|x|}{\lambda})$, where $x$ is a real number. $f'(x)=-\mbox{sign}(x)\exp(\!-\frac{|x|}{\lambda})$; therefore, $f(x)$ is a monotonic increasing function when $x<0$ and a  monotonic decreasing function when $x>0$.  Taken together, $\exp\!\left(\!\frac{c_0-s}{\lambda}\!\right)=
\exp\!\left(\!\frac{s-c_1}{\lambda}\right)$ and $s^*$ is unbiased for $s$  iff $c_0$ and $c_1$ are symmetric about $s$.

\noindent\textbf{Part b)}: When $s-c_0<c_1-s$ (both $>0$). Since $\exp\!\left(\!\frac{c_0-s}{\lambda}\!\right)\ge \exp\!\left(\!\frac{s-c_1}{\lambda}\!\right)$, then $\mu_2>s$. In the case of $\mu_1$, we have shown in Part a) that $f(x)=(|x|+\lambda)\exp(-|x|/\lambda)$ is symmetric and monotonically decreasing with $|x|$; therefore, $f(s-c_0)>f(c_1-s)$ and the numerator in Eq.~(\ref{eqn:mu1}) is $>0$. Since $\exp(\frac{s-c_1}{\lambda})<1$ and $\exp(\frac{c_0-s}{\lambda})<1$, the denominator in Eq.~(\ref{eqn:mu1}) $>0$. Taken together, $\mu_1>s$. In other words, $(\mu_1-s)(\mu_2-s)>0$. When  $s-c_0>c_1-s$,  we can prove $\mu_2<s$ and $\mu_1<s$, and therefore $(\mu_1-s)(\mu_2-s)>0$, in a similar manner as for the case of $s-c_0<c_1-s$.

\noindent \textbf{Part c)}: In terms of the magnitude of the bias, we compare
\small\begin{align*}
&\frac{\frac{s-c_0+\lambda}{2}\exp\left(\frac{c_0-s}{\lambda}\right)-\frac{c_1-s+\lambda}{2}\exp\left(\!\frac{s-c_1}{\lambda}\right)}
{1-\frac{1}{2}\exp(\frac{s-c_1}{\lambda})- \frac{1}{2}\exp(\frac{c_0-s}{\lambda})}\text{ v.s. } \frac{\lambda}{2}\left[\exp\!\left(\!\frac{c_0-s}{\lambda}\!\right)-\exp\!\left(\!\frac{s-c_1}{\lambda}\!\right)\right]\\
&\Rightarrow\!\frac{(s-c_0)\exp\left(\frac{c_0-s}{\lambda}\right)\!-\!(c_1-s)\exp\left(\!\frac{s-c_1}{\lambda}\right)\!+\!
\frac{\lambda}{2}\!\left[\exp\!\left(\!\frac{c_0-s}{\lambda}\!\right)\!-\!\exp\!\left(\!\frac{s-c_1}{\lambda}\!\right)\right]\!\!
\left[\exp(\frac{s-c_1}{\lambda})+\exp(\frac{c_0-s}{\lambda})\right]}
{1-\frac{1}{2}\exp(\frac{s-c_1}{\lambda})- \frac{1}{2}\exp(\frac{c_0-s}{\lambda})} \text{ v.s. } 0\\
&\Rightarrow\frac{(s-c_0)\exp\left(\!\frac{c_0-s}{\lambda}\!\right)\!-\!(c_1-s)\exp\left(\!\frac{s-c_1}{\lambda}\!\right)\!+\!
\frac{\lambda}{2}\left[\exp\!\left(\!2\frac{c_0-s}{\lambda}\!\right)\!-\!\exp\!\left(\!2\frac{s-c_1}{\lambda}\!\right)\right]}
{1-\frac{1}{2}\exp(\frac{s-c_1}{\lambda})- \frac{1}{2}\exp(\frac{c_0-s}{\lambda})} \text{ v.s. } 0 \\
&\textstyle\Rightarrow\!(s\!-\!c_0)\!\exp\!\left(\!\frac{c_0-s}{\lambda}\!\right)\!+\!\frac{\lambda}{2}\exp\!\left(\!2\frac{c_0-s}{\lambda}\!\right)
\!-\!\left[\!(c_1\!-\!s)\exp\!\left(\!\frac{s-c_1}{\lambda}\right)\!+\!\frac{\lambda}{2}\exp\!\left(\!2\frac{s-c_1}{\lambda}\!\right)\!\right]\text{ v.s. } 0 \text{ (the denominator $\!>\!0$.)}
\end{align*}
\normalsize
Let $f(x)=xe^{-x/\lambda}+\lambda e^{-2x/\lambda}/2$ and $x>0$, then the last equation above is to compare  $f(s-c_0)-f(c_1-s)$ v.s.  0.
The first derivative $f'(x)=-e^{-2x/\lambda}+e^{-x/\lambda}-xe^{-x/\lambda}/\lambda=e^{-x/\lambda}(1-x/\lambda-e^{-x/\lambda})$. The sign of $f'(x)$ is determined by the second term $g(x)=1-x/\lambda-e^{-x/\lambda}$ since the first term $e^{-x/\lambda}>0$. $g'(x)=\lambda(e^{-x/\lambda}-1)$. Since $x>0$, then $g'(x)<0$, implying $g(x)$ decreases monotonically with increasing $x$, and reaches the maximum as $x\rightarrow0$.  Since $g(0)=0$, so $g(x)<0$ for $x>0$. Taken together, $f'(x)=e^{-x/b}g(x)<0$. Therefore, $f(x)$ decreases monotonically with increasing $x$. When $s-c_0<c_1-s$, $f(s-c_0)>f(c_1-s)>0$ or $\mu_1-s>\mu_2-s>0$; when $s-c_0>c_1-s$, $f(s-c_0)<f(c_1-s)<0$ or $\mu_1-s<\mu_2-s<0$. In summary, $|\mu_1-s|>|\mu_2-s|$.

\section{\large{Proof of Proposition \ref{lem:error}}}\label{app:error}
For the Laplace truncated mechanism, let $A\!=\!\frac{\lambda-c_0+s}{2}\exp\left(\frac{c_0-s}{\lambda}\right)\!-\!\frac{\lambda+c_1-s}{2}\exp\left(\!\frac{s-c_1}{\lambda}\right);$
$p_1\!=\! \frac{1}{2}\exp(\frac{c_0-s}{\lambda})$; and $p_2=\frac{1}{2}\exp(\frac{s-c_1}{\lambda})$, and
$B= 2\lambda^2\!+\!s^2\!-\!
\frac{1}{2}(2\lambda^2\!-\!2\lambda c_0+c_0^2)\exp\!\left(\frac{c_0-s}{\lambda}\right)-
\frac{1}{2}(2\lambda^2\!+\!2\lambda c_1+c_1^2)\exp\!\left(\frac{s-c_1}{\lambda}\right)$. Since MSE is $\E(s^*-s)^2 = \V(s^*)+ \mbox{bias}(s^*)^2=\E(s^{*2})-(\E(s^*))^2+ \mbox{bias}^2$, thus
\begin{align*}
&\textstyle \mbox{bias}^2 =\left(\frac{A}{1-p_1-p_2}\right)^2;\;
(\E(s^*))^2= \left(s+\frac{A}{1-p_1-p_2}\right)^2;\;
\E(s^{*2}) = \frac{B}{1-p_1-p_2}\\
& \mbox{MSE} \textstyle = \frac{B}{1-p_1-p_2}-\left(s+\frac{A}{1-p_1-p_2}\right)^2+\left(\frac{A}{1-p_1-p_2}\right)^2= \frac{B-2sA}{1-p_1-p_2}-s^2; \mbox{ where } B-2sA=\\
&\textstyle2\lambda^2\!+\!s^2\!-\!
\frac{1}{2}(2\lambda^2\!-\!2\lambda c_0+c_0^2)p_1-
\frac{1}{2}(2\lambda^2\!+\!2\lambda c_1+c_1^2)p_2- s(\lambda-c_0+s)p_1+s(\lambda+c_1-s)p_2\\
=&\textstyle -s^2+2(s^2+\lambda^2)(1-p_1-p_2)
-(-\!\lambda c_0+c_0^2/2+s(\lambda-c_0))p_1
-(  \!\lambda c_1+c_1^2/2-s(\lambda+c_1))p_2
\end{align*}
Therefore, MSE
\begin{align*}
=& s^2+2\lambda^2-\frac{s^2 + (-\!\lambda c_0+\frac{c_0^2}{2}+s(\lambda-c_0))p_1+
( \!\lambda c_1+\frac{c_1^2}{2}-s(\lambda+c_1))p_2}{1-p_1-p_2}\\
= &s^2+2\lambda^2-\frac{s^2 + \!\lambda (s-c_0)p_1+(\frac{c_0^2}{2}-sc_0)p_1+
\!\lambda (c_1-s)p_2+(\frac{c_1^2}{2}-sc_1)p_2}{1-p_1-p_2}\\
= &s^2+2\lambda^2-\frac{\!\lambda\left[\!(s-c_0)p_1+\!(c_1-s)p_2\right]\!+\![(\frac{c_0^2}{2}-sc_0)p_1+(\frac{c_1^2}{2}-sc_1)p_2]+s^2}{1-p_1-p_2}\\
=&2\lambda^2-\frac{\!\lambda\!\left[(s\!-\!c_0)p_1+\!(c_1\!-\!s)p_2\right]+[(\frac{c_0^2}{2}-sc_0)p_1+(\frac{c_1^2}{2}-sc_1)p_2]+s^2(p_1+p_2)}{1-p_1-p_2}\\
=&2\lambda^2-\frac{\!\lambda\left[\!(s-c_0)p_1+\!(c_1-s)p_2\right]+[(\frac{c_0^2}{2}-sc_0+s^2)p_1+(\frac{c_1^2}{2}-sc_1+s^2)p_2]}{1-p_1-p_2}\\
=&2\lambda^2\!-\!\frac{\frac{\lambda}{2}\!\left[(s-c_0)\exp(\frac{c_0-s}{\lambda})+(c_1-s)\exp(\frac{s-c_1}{\lambda})\right]} {1-\frac{1}{2}\exp(\frac{c_0-s}{\lambda})-\frac{1}{2}\exp(\frac{s-c_1}{\lambda})}+\\
&\qquad\frac{\frac{1}{2}[(\frac{c_0^2}{2}-sc_0+s^2)\exp(\frac{c_0-s}{\lambda})+(\frac{c_1^2}{2}-sc_1+s^2)\exp(\frac{s-c_1}{\lambda})]} {1-\frac{1}{2}\exp(\frac{c_0-s}{\lambda})-\frac{1}{2}\exp(\frac{s-c_1}{\lambda})}\\
=&2\lambda^2-\\
&\boxed{\!\frac{\!\frac{\lambda}{2}\left[(s\!-\!c_0)\exp(\frac{c_0\!-\!s}{\lambda})+\!(c_1-s)\exp(\frac{s-c_1}{\lambda})\right]+
\!\frac{1}{2}\![(\frac{c_0^2}{4}\!+\!(\frac{c_0}{2}-s)^2)\exp(\frac{c_0-s}{\lambda})\!+\!(\frac{c_1^2}{4}\!+\!(\frac{c_1}{2}\!-\!s)^2)\exp(\frac{s-c_1}{\lambda})]} {1-\frac{1}{2}\exp(\frac{c_0-s}{\lambda})-\frac{1}{2}\exp(\frac{s-c_1}{\lambda})}\!\!}\\
<&2\lambda^2 \mbox{ because each term in the boxed expression is $<0$ as $c_0\le s\le c_1$}.
\end{align*}
\noindent For the BIT mechanism,
\begin{align*}
& \mbox{bias}^2 =\left(\lambda(p_1-p_2)\right)^2;\\
& (\E(s^*))^2= \left(s+\lambda(p_1-p_2)\right)^2;\;
\E(s^{*2}) =2\lambda^2+s^2-2\textstyle(\lambda^2\!-\!\lambda c_0)p_1 -2\textstyle(\lambda^2+\lambda c_1)p_2\\
\mbox{MSE} &= 2\lambda^2+s^2-2\textstyle(\lambda^2\!-\!\lambda c_0)p_1 -2\textstyle(\lambda^2+\lambda c_1)p_2- \left(s+\lambda(p_1-p_2)\right)^2+\left(\lambda(p_1-p_2)\right)^2\\
&= 2\lambda^2(1-p_1-p_2)+2\lambda((c_0-s)p_1+(s-c_1)p_2)\\
&= 2\lambda^2+ \boxed{\left(-2\lambda^2p_1-2\lambda^2p_2+2\lambda((c_0-s)p_1+(s-c_1)p_2)\right)}\\
&<2\lambda^2\mbox{ because each term in the boxed expression is $<0$ as $c_0\le s\le c_1$}.
\end{align*}
Since $\lambda = \delta_s/\epsilon$; if $\delta_s\propto n^{-k}$ (for $k>0$), then MSE $=O(\lambda^2)  = O(n^{-2k})$ for a given $\epsilon$ in both the truncated and BIT mechanisms.

With the established upper bound $2\lambda^2$ as above, it is obvious that MSE $\rightarrow 0$ as $\lambda \rightarrow 0$.

\section{\large{Proof of Proposition \ref{prop:asymp}}}\label{app:lem}
The proof utilizes the following lemma.

\noindent \textbf{Lemma A.1}: If 1) an estimator $\hat{\theta}$ is asymptotically unbiased for $\theta$ ($\E(\hat{\theta})\rightarrow\theta$ as $n\rightarrow\infty$), and 2) there exists  a $k\ge0$ such that $\int_-k<\infty$ and $|\E(\hat{\theta}|\theta)f(\theta|\beta)|\leqq k$ for all $n$, where $f(\theta|\beta)$ is a probability density function, 
then $\E(\E(\hat{\theta}|\theta)|\beta) \rightarrow\E(\theta|\beta)$ as  $n\rightarrow\infty$. 

\noindent Proof:  $\E(\hat{\theta})\rightarrow\theta$ as $n\rightarrow \infty$ and $\E(\hat{\theta})f(\theta|\beta)\rightarrow\theta f(\theta|\beta)$ as $n\rightarrow \infty$.  With condition 2) and Theorem 2 from Cunningham (1967), we have $\int\E(\hat{\theta})f(\theta|\beta)d\theta \rightarrow\int \theta f(\theta|\beta)d\theta=\E(\theta|\beta)$ as $n\rightarrow\infty$.

\noindent \textbf{Part a)}: By the  the law of total expectation, $\E(\s^*|\boldsymbol{\theta})=\E[\E(\s^*|\s)|\boldsymbol{\theta}]$. Since $\E(\s^*|\s)\rightarrow\s$, \\ $\E[\E(\s^*|\s)|\boldsymbol{\theta}]\!\rightarrow\!\E[\s|\boldsymbol{\theta}]$ by Proposition A.1. Since $ \E[\s|\boldsymbol{\theta}]\rightarrow\boldsymbol{\theta}$, then $\E(\s^*|\boldsymbol{\theta})]\rightarrow \boldsymbol{\theta}$.

\noindent  \textbf{Part b)}: By the  the law of total variance, $\V(\s^*|\boldsymbol{\theta})=\V[\E(\s^*|\s)|\boldsymbol{\theta}]+\E[\V(\s^*|\s)|\boldsymbol{\theta}]$. Since $\s^*\xrightarrow{p}{}\s$, $\V(\s^*|\s)\rightarrow 0$ as $n\rightarrow0$. By Proposition A.1, $\E[\V(\s^*|\s)|\boldsymbol{\theta}]\rightarrow0$ as $n\rightarrow0$.  Since $\s^*\xrightarrow{p}{}\s$, then $\E(\s^*|\s)\rightarrow 0$ as $n\rightarrow0$. By Proposition A.1, $\V[\E(\s^*|\s)|\boldsymbol{\theta}]\rightarrow0$ as $n\rightarrow0$. By part b),   $\E(\s^*|\s)\rightarrow \s$, and  $\E(\s|\boldsymbol{\theta})\rightarrow\boldsymbol{\theta}$, then $\E(\s^*|\boldsymbol{\theta})\rightarrow\boldsymbol{\theta}$ as $n\rightarrow0$. All taken together,  $\s^*\xrightarrow{p}{}\boldsymbol{\theta}$.

\bibliographystyle{apalike}

\newpage
\setstretch{1.25}
\begin{center}
\large{\textbf{Supplemental Materials to\\
``Statistical Properties of Sanitized Results from Differentially Private Laplace Mechanism with Univariate Bounding Constraints''
}}
\end{center}

\normalsize
\setstretch{1.0}
There are two ways defining two data sets differing by one record $\Delta(\x,\x')=1$. In the first definition,  referred to as ``Def 1'' below, the two data sets have the same sample size $n$, but one and only one record  differs in at least one attributes; a substitution would make the two data sets identical. In the second definition, referred to as ``Def 2'' below, one data set has one more record that the other, so the sample sizes differ by 1 (one is $n$ and the other $n-1$), and a deletion (or an insertion) would make the two data sets identical.  We calculated the $l_1$ GS for some common statistics below in both ways, and the results turned out to be the same, as shown below.   Without loss of generality (WLOS), we assume it is the first observation that differs in data sets $\x$ and $\x'$ in Def 1, and $\x'$ does not have the first row ($x_1$) compared to $\x$ in Def 2 in the following calculation. Def 2 is more intuitive from the perspective of interpreting global sensitivity GS and DP, but the calculation of GS under Def 1 in general is much simpler (when $\x$ and $\x'$ are of the same size) analytically than that under Def 2. In most cases, the two definitions lead to the same GS (such as mean, variance, and covariance); In the other two statistics (pooled variance and pooled covariance across multiple groups), the GS calculated under the two definitions are different, but the discrepancy  between the two in terms of its impact on the sanitized results usually diminishes when $n$ gets large.

Sections \ref{app:GSmean} to \ref{app:GScovp} presents the global sensitivity for mean, variance, and covariance; and Section \ref{app:simulation} provides additional simulation results in the second simulation study on releasing a vector of proportions (a histogram). 

\section{\texorpdfstring{$l_1$}{}-GS of sample mean and proportion}\label{app:GSmean}
Denote  by  $\delta_{\bar{x}}$ the GS of a sample mean of variable $x$ that  is globally bounded in $[c_0,c_1]$, then
\begin{flalign*}
\mbox{\textbf{Def 1: }}\delta_{\bar{x}}&=\sup_{\x,\x':\Delta(\x,\x')=1}\left|\frac{1}{n}\sum^n_{i=1}x_i-\frac{1}{n}\sum^{n}_{i=1}x'_i\right|
=\sup_{\x,\x':\Delta(\x,\x')=1}n^{-1}\left|x_1-x'_1\right|=\textcolor{red}{n^{-1}(c_1-c_0)}. \;\blacksquare\\
\mbox{\textbf{Def 2: }}\delta_{\bar{x}}& =\sup_{\x,\x':\Delta(\x,\x')=1}\textstyle\left|\frac{1}{n}\sum^n_{i=1}x_i-\frac{1}{n-1}\sum^{n}_{i=2}x_i\right|\\
&=\sup_{\x,\x':\Delta(\x,\x')=1}\left|n^{-1}((n-1)\bar{x}_{-}+x_n)-\bar{x}_{-}\right| =n^{-1}\sup_{\x,\x':\Delta(x,x')=1}\left|x_n-\bar{x}_{-}\right|,\\
\mbox{where } &\mbox{ $\bar{x}_{-}=\textstyle(n-1)^{-1}\sum^n_{i=2}x_i$. The maximum possible value of $|x_n-\bar{x}|$ is $c_1-c_0$}\\
\mbox{across } & \mbox{all possible data sets $\x$ and all possible ways leading to $\Delta(\x,\x')=1$.} \\
&\mbox{ Therefore,  \textcolor{red}{$\delta_{\bar{x}}= n^{-1}(c_1-c_0)$}.\hspace{9cm}\qedsymbol}
\end{flalign*}

A sample proportion can be viewed as a special case of  a sample mean with the mean operated on the indicator function with $c_1=1$ and $c_0=0$.  Therefore,  $\delta_1$ of a single proportion is $n^{-1}$.  In addition to releasing a single proportion, in many practical cases, it is of interest to release a whole histogram $\mathbf{H}$ or a whole vector of proportions $\mathbf{p}$, the GS of which differ between the two definitions of $\Delta(\x,\x')$. Specifically, in Def 1, where $n$ is the same between $\x$ and $\x'$, $\delta_1=2$ and  $\delta_1=2n^{-1}$ for $\mathbf{H}$ and $\mathbf{p}$, respectively; in Def 2, $\delta_1=1$ and  $\delta_1=n^{-1}$.

\section{\texorpdfstring{$l_1$}{}-GS of sample variance}\label{app:GSvar}
Denote the sample variances of $\x$ and $\x'$ by $s^2$ and $s'^2$, respectively, and the global bounds by  $[c_0,c_1]$. The $l_1$ GS $\delta_{s^2}=\sup_{\x,\x':\Delta(\x,\x')=1}|s^2-s'^2|$, where
\baselineskip=0pt
\begin{align}
\mbox{\textbf{Def 1}: }& s^2-s'^2
=(n-1)^{-1}\left(\textstyle \sum^n_{i=1}(x_i^2-x_i^{'2})- n\left(\bar{x}^2-\bar{x'}^2\right)\right)\notag\\
&=(n-1)^{-1}\left(x_1^2-x_1^{'2}- n^{-1}(x_1-x'_1)(2\textstyle \sum_{i=2}^nx_i+x_1+x'_1)\right)\notag\\
&=(n-1)^{-1}(x_1-x'_1) \left(x_1+x'_1- n^{-1}(2\textstyle \sum_{i=2}^nx_i+x_1+x_1') \right)\notag\\
&=(n-1)^{-1}(x_1-x'_1) \left((1-n^{-1})x_1+(1-n^{-1})x'_1 -2n^{-1}\textstyle \sum_{i=2}^nx_i\right)\notag\\
&=n^{-1}(x^2_1-x_1^{'2})- 2n^{-1}(x_1-x'_1) \bar{x}_{-},\mbox{ where }\bar{x}_{-}=(n-1)^{-1}\textstyle \sum_{i=2}^nx_i\notag\\
&=n^{-1}(x^2_1-2x_1\bar{x}_{-}+\bar{x}_{-}^2 -x_1^{'2}+2x'_1\bar{x}_{-}-\bar{x}_{-}^2)\notag\\
&=n^{-1}(x_1-\bar{x}_{-})^2- n^{-1}(x'_1-\bar{x}_{-})^2.\label{eq:var}
\end{align}
where  $\bar{x}_{-}=\textstyle(n-1)^{-1}\sum^n_{i=2}x_i$. Since both terms in Eq (\ref{eq:var}) are $\ge0$, $|s^2-s'^2|$ is maximized when $(x_1-\bar{x}_{-})^2$ reaches its maximum $(c_1-c_0)^2$ and $(x'_1-\bar{x}_{-})^2=0$, or when $(x_1-\bar{x}_{-})^2=0$ and $(x'_1-\bar{x}_{-})^2$  reaches its maximum $(c_1-c_0)^2$. Therefore,
$$\textcolor{red}{\delta_{s^2}=\sup_{\x,\x':\Delta(\x,\x')=1}|s^2-s'^2|= n^{-1}(c_1-c_0)^2}.\quad\blacksquare$$
\begin{align*}
\mbox{\textbf{Def 2: }}& s^2-s'^2
=\frac{x_1^2+\sum^n_{i=2}x_i^2-n^{-1}((n-1)\bar{x}_{-}+x_1)^2}{n-1}-\frac{\sum^n_{i=2}x_i^2-(n-1)\bar{x}_{-}^2}{n-2}\\
&=[(n-1)(n-2)]^{-1}\left[(n-2)x_1^2+(n-2)\textstyle\sum^n_{i=2}x_i^2\right.\\
&\textstyle\left.-(n-2)n^{-1}\left((n-1)^2\bar{x}_{-}^2+2(n-1)x_1\bar{x}_{-}+x_1^2\right)-(n-1)\sum^n_{i=2}x_i^2+(n-1)^2\bar{x}_{-}^2\right]\\
&=\frac{-\sum^n_{i=2}x_i^2+(n-2)(1-n^{-1})x_1^2+2n^{-1}(n-1)^2\bar{x}_{-}^2-2n^{-1}(n-1)(n-2)\bar{x}_{-}x_1}{(n-1)(n-2)}\\
&=-\frac{\sum^n_{i=2}x_i^2}{(n-1)(n-2)}+\frac{x_1^2}{n}+2\frac{(n-1)}{n(n-2)}\bar{x}_{-}^2-2\frac{\bar{x}_{-}x_1}{n}\\
&=\frac{1}{n}\left[x_1^2-2\bar{x}_{-}x_1+\bar{x}_{-}^2\right]-\frac{\bar{x}_{-}^2}{n}+\frac{2(n-1)\bar{x}_{-}^2}{n(n-2)}-\frac{\sum^n_{i=2}x_i^2}{(n-1)(n-2)}\\
&=\frac{1}{n}(x_1-\bar{x}_{-})^2+\frac{\bar{x}_{-}^2}{n-2}-\frac{\sum^n_{i=2}x_i^2}{(n-1)(n-2)}\\
&=\frac{1}{n}(x_1-\bar{x}_{-})^2-\frac{\sum^n_{i=2}x_i^2-(n-1)\bar{x}_{-}^2}{(n-1)(n-2)}
= n^{-1}(x_1-\bar{x}_{-})^2-(n-1)^{-1}s'^2.
\end{align*}
$|s^2-s'^2|$  is maximized  in either of the following two cases, whichever is larger: 1) when $(x_1-\bar{x}_{-})^2=0$ and  $s'^2$ reaches maximum, and 2) when $s'^2=0 $ and $(x_1-\bar{x}_{-})^2$  reaches maximum.  Case 1) occurs when $x_1=\bar{x}_{-}$ and  $s'^2=(c_1-c_0)^2\frac{n-1}{4(n-2)}$, the maximum possible value of the sample variance with sample size $(n-1)$ (Shiffler and Harsha, 1980), leading to
$\sup_{\x,\x':\Delta(\x,\x')=1}|s^2-s'^2|=(c_1-c_0)^2/(4n-8)$.  Case 2) is realized when $(x_1-\bar{x}_{-})^2=(c_1-c_0)^2$, and $s'^2=0$ (when all $x_i=c_0$ for $i=2,\ldots,n$ if $x_1=c_1$, or when all $x_i=c_1$ for $i=2,\ldots,n$ if $x_1=c_0$), leading to $\sup_{\x,\x':\Delta(\x,\x')=1x}|s^2-s'^2|=(c_1-c_0)^2/n.$ Since Case 2 is larger (for $n>2$), then $$\textcolor{red}{\delta_{s^2}= n^{-1}(c_1-c_0)^2}.\quad\blacksquare$$

\section{\large{\texorpdfstring{$l_1$}{}-GS of sample covariance}}\label{app:GScov}
Denote the sample covariance between $x_1$ and $x_2$ in $\x$ by $s_{12}$ and that in $\x'$ by $s'_{12}$ respectively.   Denote the global bounds of $x_1$ and $x_2$ by  $[c_{10},c_{11}]$ and $[c_{20},c_{21}]$, respectively.
\baselineskip=0pt
\begin{align}
\mbox{\textbf{Def 1}: }  & s_{12}^2-s_{12}'^2
=(n-1)^{-1}[\textstyle \sum^n_{i=1}x_{i1}x_{i2}- n\bar{x}_1\bar{x}_2- \sum^n_{i=1}x'_{i1}x'_{i2}+n\bar{x}'_1\bar{x}'_2]\notag\\
&=(n-1)^{-1}[x_{11}x_{12}-x'_{11}x'_{21}- n^{-1}(x_{11}+(n-1)\bar{x}_{1-})(x_{12}+(n-1)\bar{x}_{2-})\notag\\
&\qquad\qquad\qquad+ n^{-1} (x'_{11}+(n-1)\bar{x}_{1-})(x'_{21}+(n-1)\bar{x}_{2-})]\notag\\
&=n^{-1}[ x_{11}x_{12}-x'_{11}x'_{21}- x_{11}\bar{x}_{2-}-\bar{x}_{1-}x_{12}+x'_{11}\bar{x}_{2-}+\bar{x}_{1-}x'_{21}]\notag\\
&=n^{-1}[ x_{11}\underbrace{(x_{12}-\bar{x}_{2-})}_{\mbox{ term 1}}+\bar{x}_{1-}\underbrace{(x'_{21}-x_{12})}_{\mbox{ term 2}}+ x'_{11}\underbrace{(\bar{x}_{2-}-x'_{21})}_{\mbox{ term 3}}],\label{eqn:def1s12}
\end{align}
where $\bar{x}_{k-}=(n-1)^{-1}\sum_{i=2}^n x_{ki}$ ($k=1,2$).  WLOS, assume that $x_{2-}\le x_{12}\le x_{12}'$, then term 1 and term 2 $>0$, term 3 $<0$. Eq (\ref{eqn:def1s12}) is maximized when  $x_{11}=\bar{x}_{1-} = c_{11}$  and $\bar{x}_{1-}=c_{10}$, and Eq (\ref{eqn:def1s12}) is then written as $n^{-1}[ c_{11}(x_{12}-\bar{x}_{2-}) + c_{11}(x'_{21}-x_{12}) +c_{10}(\bar{x}_{2-}-x'_{21})]=
(c_{11}-c_{10})(x'_{21}-\bar{x}_{2-})$, which reaches its maximum when $x'_{21}-\bar{x}_{2-}= c_{21}-c_{20}$. Therefore,
\baselineskip=0pt
$$\textcolor{red}{\delta_{s_{12}}=\sup_{\x,\x':\Delta(\x,\x')=1}|s_{12}-s'_{12}|= n^{-1}(c_{11}-c_{10}) (c_{21}-c_{20})}.\quad\blacksquare$$
\baselineskip=0pt
\textbf{Def 2}:  Denote the sample mean of $\x'$ with one less observation by $\bar{\x}$. The GS of the covariance is $\delta_{s_{12}}=\sup_{\x,\x':\Delta(\x,\x')=1}|s_{12}-s'_{12}|$, where
\begin{align}
\Delta & s_{12}-s'_{12}=\frac{\sum^n_{i=2}x_{i1}x_{i2}+x_{11}x_{12}-n^{-1}[((n-1)\bar{x}_{1-}+x_{11})((n-1)\bar{x}_{2-}+x_{12})]}{n-1}\notag\\
&\hspace{2cm}-\frac{\sum^n_{i=2}x_{i1}x_{i2}-(n-1)\bar{x}_{1-}\bar{x}_{2-}}{n-2}\notag\\
=&\frac{1}{(n-1)(n-2)}\!\left[(n-2)\textstyle\!\sum^n_{i=2}x_{i1}x_{i2}\!+\!(n-2)x_{11}x_{12}\!-\!(n-1)\!\sum^n_{i=2}x_{i1}x_{i2}\!+\!(n-1)^2\bar{x}_{1-}\bar{x}_{2-}\right.\notag\\
&\hspace{2cm}-\left.(n-2)n^{-1}\left((n-1)^2\bar{x}_{1-}\bar{x}_{2-}+(n-1)\bar{x}_{1-}x_{12}+(n-1)x_{11}\bar{x}_{2-}+x_{11}x_{12}\right)\right]\notag\\
=&\frac{1}{(n-1)(n-2)}\left[-\textstyle\sum^n_{i=2}x_{i1}x_{i2}+n^{-1}(n-1)(n-2)x_{11}x_{12}+2n^{-1}(n-1)^2\bar{x}_{1-}\bar{x}_{2-}\right.\notag\\
&\hspace{2cm}\left.-n^{-1}(n-1)(n-2)(\bar{x}_{1-}x_{12}+x_{11}\bar{x}_{2-})\right]\notag\\
=&-\frac{\textstyle\sum^n_{i=2}x_{i1}x_{i2}}{(n-1)(n-2)}+\frac{x_{11}x_{12}}{n}+2\frac{(n-1)}{n(n-2)}\bar{x}_{1-}\bar{x}_{2-}-\frac{\bar{x}_{1-}x_{12}+x_{11}\bar{x}_{2-}}{n}\notag\\
=&\frac{1}{n}\left[x_{11}x_{12}-\bar{x}_{1-}x_{12}-x_{11}\bar{x}_{2-}+\bar{x}_{1-}\bar{x}_{2-}\right]- \frac{\bar{x}_{1-}\bar{x}_{2-}}{n}-\frac{\sum^n_{i=2}x_{i1}x_{i2}}{(n-1)(n-2)}+2\frac{(n-1)}{n(n-2)}\bar{x}_{1-}\bar{x}_{2-}\notag\\
=&\frac{1}{n}(x_{11}-\bar{x}_{1-})(x_{12}-\bar{x}_{2-})+\frac{\bar{x}_{1-}\bar{x}_{2-}}{n-2}-\frac{\sum^n_{i=2}x_{i1}x_{i2}}{(n-1)(n-2)}\notag\\
=&n^{-1}(x_{11}-\bar{x}_{1-})(x_{12}-\bar{x}_{2-})-(n-1)^{-1}s'_{12}\label{eq:def2s12}
\end{align}
Both terms $(x_{11}-\bar{x}_{1-})(x_{12}-\bar{x}_{2-})$ and $s'_{12}$ in Eq (\ref{eq:def2s12}) can be $>0$ or $<0$ and depend on $\bar{x}_{1-}$ and $\bar{x}_{2-}$, making the determination of the maximum of $|s_{12}-s'_{12}|$ complicated.  Rewrite Eq (\ref{eq:def2s12}) as
\begin{align*}
\Delta=&n^{-1}\textstyle\left(x_{11}-(n-1)^{-1}\sum_{i=2}^nx_{i1}\right)\left(x_{12}-(n-1)^{-1}\sum_{i=2}^nx_{i2}\right)+\\
&(n-2)^{-1}(n-1)^{-2}\textstyle\sum_{i=2}^n x_{i1}\sum_{i=2}^nx_{i2} -(n-2)^{-1}(n-1)^{-1}\sum^n_{i=2}x_{i1}x_{i2},
\end{align*}
which suggests $\Delta$ is a linear function of $x_{i1}$ and $x_{i2}$ for all $i=1,\ldots,n$, indicating that the maximum of $|s_{12}-s'_{12}|$ occurs at the corners of  the vector $(x_{11},\ldots,x_{n1},x_{12},\ldots,x_{n2})$.  To simplify the algebra, we work with transformed $x_{i1}$ and $x_{i2}$, that is,  $x_{i1}=(c_{11}-c_{10})^{-1}(x_{i1}-c_{10})$ and $x_{i2}=(c_{21}-c_{20})^{-1}(x_{i2}-c_{20})$.  After the transformation, $x_{i1}$ and $x_{i2}$ are both bounded within $[0,1]$.  The goal is to determine between 0 and 1 at each of the $2n$ position that lead to the maximum  $|s_{12}-s'_{12}|$.   The maximum $\Delta$ on the original scale can be obtained by scaling the maximum $\Delta$ on the $[0,1]\times[0,1]$ scale with $(c_{11}-c_{10})(c_{21}-c_{20})$.

Let $k_1=\#\{x_{i1}=1\}, k_2=\#\{x_{i2}=1\}$ and $k_3=\sum_i{x_{i1}x_{i2}}$ for $i=2,\ldots,n$. Thus $\bar{x}_1=(n-1)^{-1}k_1$, and $\bar{x}_2=(n-1)^{-1}k_2$.
It is easy to see $k_3\in[\max(0, k_1+k_2-(n-1)), \min(k_1,k_2)]$. WLOS, assume $k_2\le k_1$, thus $0\le k_1\le n-1, 0\le k_2\le k_1, \max(0, k_1+k_2-(n-1))\le k_3\le k_2$. Among the three, $k_1$ vary from 0 to $n-1$, while the range of $k_2$ depends on $k_1$, and that of $k_3$ depends on  $k_1$ and $k_2$.
\begin{enumerate}
\item If $(x_{11},x_{12})=(0,0)$, then
$$s_{12}-s'_{12}=\frac{k_1k_2}{n(n-1)^2}-\frac{k_3-(n-1)^{-1}k_1k_2}{(n-1)(n-2)}=\frac{2k_1k_2}{n(n-1)(n-2)}-\frac{k_3}{(n-1)(n-2)}$$
$s_{12}-s'_{12}$ is linear in $k_1,k_2$ and $k_3$, thus the minimum/maximum occurs at corners of $(k_1,k_2,k_3)$.
\begin{itemize}
\item If $k_1=0$, then $k_2=0, k_3=0$, and  \textcolor{blue}{$s_{12}-s'_{12}=0$}
\item If $k_1=n-1$, then $k_2\in[0,n-1], k_3=k_2$, and  $s_{12}-s'_{12}=\frac{k_2}{n(n-1)}$,  thus the \textcolor{blue}{maximum of $|s_{12}-s'_{12}|$ is $n^{-1}$} when $k_1=k_2=k_3=n-1$
\end{itemize}
\item If $(x_{11},x_{12})=(0,1)$, then
\begin{align*}
s_{12}-s'_{12}&=-\frac{k_1(n-1-k_2)}{n(n-1)^2}-\frac{k_3-(n-1)^{-1}k_1k_2}{(n-1)(n-2)}\\
&=\frac{2k_1k_2}{n(n-1)(n-2)}-\frac{k_3}{(n-1)(n-2)}-\frac{k_1}{n(n-1)}
\end{align*}
$s_{12}-s'_{12}$ is linear in $k_1,k_2$ and $k_3$, thus the minimum/maximum occurs at corners of $(k_1,k_2,k_3)$.
\begin{itemize}
\item If $k_1=0$, then $k_2=0, k_3=0$, and  \textcolor{blue}{$s_{12}-s'_{12}=0$}
\item If $k_1=n-1$, then $k_2\in[0,n-1], k_3=k_2$, and  $s_{12}-s'_{12}=\frac{k_2}{n(n-1)}-\frac{1}{n}$.  when $k_2=0$, $s_{12}-s'_{12}=-n^{-1}$, when $k_2=(n-1)$, $s_{12}-s'_{12}=0$, thus \textcolor{blue}{the maximum of $|s_{12}-s'_{12}|$ is $n^{-1} $} when $k_1=n-1, k_2=k_3=0$
\end{itemize}
\item If $(x_{11},x_{12})=(1,0)$, then
 \begin{align*}
s_{12}-s'_{12}&=-\frac{k_2(n-1-k_1)}{n(n-1)^2}-\frac{k_3-(n-1)^{-1}k_1k_2}{(n-1)(n-2)}\\
&=\frac{2k_1k_2}{n(n-1)(n-2)}-\frac{k_3}{(n-1)(n-2)}-\frac{k_2}{n(n-1)}
\end{align*}
$s_{12}-s'_{12}$ is linear in $k_1,k_2$ and $k_3$, thus the minimum/maximum occurs at corners of $(k_1,k_2,k_3)$.
\begin{itemize}
\item If $k_1=0$, then $k_2=0, k_3=0$, and  \textcolor{blue}{$s_{12}-s'_{12}=0$}
\item If $k_1=n-1$, then $k_2\in[0,n-1], k_3=k_2$, and  \textcolor{blue}{$s_{12}-s'_{12}=\frac{k_2}{n(n-1)}-\frac{k_2}{n(n-1)}=0$}.
\end{itemize}
\item If $(x_{11},x_{12})=(1,1)$, then
\begin{align*}
s_{12}-s'_{12}&=-\frac{(n-1-k_1)(n-1-k_2)}{n(n-1)^2}-\frac{k_3-(n-1)^{-1}k_1k_2}{(n-1)(n-2)}\\
&= \frac{1}{n}+\frac{2k_1k_2}{n(n-1)(n-2)}-\frac{k_3}{(n-1)(n-2)}-\frac{k_1}{n(n-1)}-\frac{k_2}{n(n-1)}
\end{align*}
$s_{12}-s'_{12}$ is linear in $k_1,k_2$ and $k_3$, thus the minimum/maximum occurs at corners of $(k_1,k_2,k_3)$.
\begin{itemize}
\item If $k_1=0$, then $k_2=0, k_3=0$, and  $s_{12}-s'_{12}=n^{-1}$, thus\textcolor{blue}{ the maximum of $|s_{12}-s'_{12}|$ is $n^{-1}$} when $k_1=k_2=k_3=0$
\item If $k_1=n-1$, then $k_2\in[0,n-1], k_3=k_2$, and  \textcolor{blue}{$s_{12}-s'_{12}=0$.}
\end{itemize}
\end{enumerate}
Note that the above results are obtained under the assumption $k_2\le k_1$. The same sets of results are obtained by changing the assumption to $k_1\le k_1$, and flipping the labels the two variables in each of the 4 cases above. In summary, on the transformed scale $[0,1]\times[0,1]$,  \textcolor{blue}{the maximum of $|s_{12}-s'_{12}|$ is $n^{-1}$}, that occurs if any one of the following conditions holds: 1) $x_{11}=x_{12}=0$, and $x_{i1}=1, x_{i2}=1$ for all $i=2,\ldots,n$; 2) $x_{11}=x_{12}=1$, and $x_{i1}=0, x_{i2}=0$ for all $i=2,\ldots,n$; 3) $x_{11}=0, x_{12}=1$, and $x_{i1}=1$ and $x_{i2}=0$ for all $i=2,\ldots,n$;  4) $x_{11}=1, x_{12}=0$, and $x_{i1}=0$ and $x_{i2}=1$ for all $i=2,\ldots,n$; Transforming back to the original scale $[c_{10},c_{11}]\times[c_{20},c_{21}]$, we have
$$\textcolor{red}{\sup_{\x,\x':\Delta(\x,\x')=1}|s_{12}-s'_{12}|=n^{-1}(c_{11}-c_{10})(c_{21}-c_{20})},$$
which occurs if any one of the following conditions holds: 1) $x_{11}=c_{10}$, $x_{12}=c_{20}$, and $x_{i1}=c_{11}, x_{i2}=c_{21}$ for all $i=2,\ldots,n$; 2) $x_{i1}=c_{11}, x_{i2}=c_{21}$, and $x_{i1}=c_{10}, x_{i2}=c_{20}$  for all $i=2,\ldots,n$.; 3) $x_{11}=c_{10}, x_{12}=c_{21}$, and $x_{i1}=c_{11}$ and $x_{i2}=c_{20}$ for all $i=2,\ldots,n$;  and 4) $x_{11}=c_{11}, x_{12}=c_{20}$, and $x_{i1}=c_{10}$ and $x_{i2}=c_{21}$ for all $i=2,\ldots,n$;
\section{\large{\texorpdfstring{$l_1$}{}-GS  pooled sample variance}}\label{app:GSvarp}
Denote the number of cells of $J$ and  $n_j$ is the number of observations in cell $j$ (for $j=1,\ldots,J$). Assume each cell contributes to the pooled variance $s_p^2$; in other words, there are at least 2 observations in each cell ($n_j\ge2$ for $j=1,\ldots,J$). Denote the total sample size by $n=\sum_{j=1}^Jn_j$, then
\baselineskip=0pt
$$s_p^2=(n-J)^{-1}\left(\textstyle\sum_{j=1}^J \sum_{i=1}^{n_j} (x_{ij}-\bar{x}_j)^2\right),$$
where  $x_{ij}$ is the $i$-th observation in cell $j$, and $\bar{x}_j$ is the mean of cell $j$.

\textbf{Def 1}: WLOS, suppose it is the first observation in cell $j=1$ that differs between data $\x$ and $\x'$, then
\baselineskip=0pt
\begin{align*}
\Delta =s_p^2-s^{'2}_p &= (n-J)^{-1}\left(\textstyle\sum_{j=1}^J \sum_{i=1}^{n_j} (x_{ij}-\bar{x}_j)^2\right)-(n-J)^{-1}\left(\textstyle\sum_{j=1}^J \sum_{i=1}^{n_j} (x'_{ij}-\bar{x'}_j)^2\right)\\
&= (n-J)^{-1}(n_1-1)\underbrace{(n_1-1)^{-1}\left(\textstyle\sum_{i=1}^{n_1} (x_{i1}-\bar{x}_1)^2-\sum_{i=1}^{n_1} (x_{i1}-\bar{x}'_1)^2\right)}_{\text{term 1}}.
\end{align*}
Term 1 is the difference in the variance in cell 1 between $\x$ and $\x$, and its maximum is $n^{-1}_1(c_1-c_0)^2$ per the results in Section \ref{app:GSvar}. Therefore, max$|\Delta|=(n-J)^{-1}(c_1-c_0)^2(1-n^{-1}_1)$, which reaches maximum if $n_1$ is the largest among $(n_1,\ldots,n_J)$. All taken together,  the GS of $s_p^2$ is
\baselineskip=0pt
\begin{align*}
&\textcolor{red}{\delta_{s_p^2} = (c_1-c_0)^2 (n-J)^{-1}\left(1-n_{\text{max}}^{-1}\right)},\mbox{ which can be approximated by }\\
&\textcolor{red}{\delta_{s_p^2} = (c_1-c_0)^2 (n-J)^{-1}}
\end{align*}
if $n_{\text{max}}$ is large or when $n_{\text{max}}$ itself cannot be released without sanitization. $\blacksquare$

\textbf{Def 2}:  WLOS, suppose the first observation in cell $1$ is removed in $\x'$ compared to $\x$, then
$$s_{p-}^{2}=(n-1-J)^{-1}\left(\textstyle\sum_{j=2}^J \sum_{i=1}^{n_j} (x_{ij}-\bar{x}_j)^2+ \sum_{i=2}^{n_1} (x_{i1}-\bar{x}_{1-})^2\right),$$
where $\bar{x}_{1-}$ is the mean of cell 1 without the first observation. Let \small
\begin{align*}
&\Delta =s_p^2-s^2_{p-}=\frac{(n-1-J)\left(\sum_{j=2}^J\sum_{i=1}^{n_j} (x_{ij}-\bar{x}_j)^2+\sum_{i=1}^{n_1} (x_{i1}-\bar{x}_1)^2\right)}{(n-J)(n-1-J)}\\
&\qquad \qquad\qquad \qquad  -\frac{(n-J)\left(\sum_{j=2}^J \sum_{i=1}^{n_j} (x_{ij}-\bar{x}_j)^2+ \sum_{i=2}^{n_1} (x_{i1}-\bar{x}_{1-})^2\right)}{(n-J)(n-1-J)}\\
&=\frac{(n-1-J)\sum_{i=1}^{n_1} (x_{i1}-\bar{x}_1)^2-(n-J)\sum_{i=2}^{n_1} (x_{i1}-\bar{x}_{1-})^2-\sum_{j=2}^J\sum_{i=1}^{n_j} (x_{ij}-\bar{x}_j)^2}{(n-J)(n-1-J)}\\
&=\frac{(n-1-J)\sum_{i=1}^{n_1} (x_{i1}-\bar{x}_1)^2-(n-J)\sum_{i=2}^{n_1} (x_{i1}-\bar{x}_{1-})^2+\sum_{i=2}^{n_1} (x_{i1}-\bar{x}_{1-})^2}{(n-J)(n-1-J)}\\
&\quad-\frac{\sum_{i=2}^{n_1} (x_{i1}-\bar{x}_{1-})^2+\sum_{j=2}^J\sum_{i=1}^{n_j} (x_{ij}-\bar{x}_j)^2}{(n-J)(n-1-J)}\\
&=\frac{\left(\sum_{i=1}^{n_1} (x_{i1}-\bar{x}_1)^2-\sum_{i=2}^{n_1} (x_{i1}-\bar{x}_{1-})^2\right)}{(n-J)}-\frac{s^2_{p-}}{n-J}\\
&=\frac{\sum_{i=1}^{n_1} x_{i1}^2-\frac{((n_1-1)\bar{x}_{1-}+x_{11})^2}{n_1}-\sum_{i=2}^{n_1}x_{i1}^2+(n_1-1)\bar{x}_{1-}^2-s_{p-}^2}{n-J}\\
&=\frac{ (1-n_1^{-1})x_{11}^2+(1-n_1^{-1})\left(\bar{x}_{1-}^2-2\bar{x}_1x_{11}\right)-s_{p-}^2}{n-J}
= \frac{ (1-n_1^{-1})(x_{11}-\bar{x}_{1-})^2-s_{p-}^2}{n-J}
\end{align*}
\normalsize Since $(x_{11}-\bar{x}_{1-})^2\ge0$ and $s_{p-}^2\ge0$, the maximum of $|\Delta|$ takes the larger value of the two:  max\{$(x_{11}-\bar{x}_1)^2$\} (case 1; $s_{p-}^2=0$ in this case), and  max\{$s_{p-}^2$\} (case 2; $(x_{11}-\bar{x}_{1-})^2=0$.
\begin{itemize}
\item case 1: $(x_{11}-\bar{x}_1)^2=(c_1-c_0)^2$ when $(x_{11},\bar{x}_{1-})=(c_0,c_1)$ or $(c_1,c_0)$. If $s_{p-}^2=0$, then the observations in a cell $j>1$ are the same, and all the observations in cell 1 (without the first case ) are the same (if $\bar{x}_{1-}=c_0$, then $x_{i1}\equiv c_0$ for $i=2,\ldots,n_1$; if $\bar{x}_{1-}=c_1$, then $x_{i1}\equiv c_1$ for $i=2,\ldots,n_1$). Therefore, max$|\Delta|=(1-n_1^{-1})(c_1-c_0)^2/(n-J)$, which is again is maximized when $n_1$ is the maximum among all cell sizes. That is, max$|\Delta |=(1-n_{\text{max}}^{-1})(c_1-c_0)^2/(n-J)$.
\item case 2: $s_{p-}^{2}$ reaches its maximum if the sum of squares of $x$ in each cell reaches its maximum, which is $(c_1-c_0)^2n_j/4$ in cell $j\ge2$ and  $(c_1-c_0)^2(n_1-1)/4$ in cell 1.  Therefore, max$\{s_{p-}^2\}=\frac{(c_1-c_0)^2}{4}\frac{(n_1-1)+\sum_{j=2}^J n_j}{n-1-J}= \frac{(n-1)(c_1-c_0)^2}{4(n-1-J)} $ and max$|\Delta|=(c_1-c_0)^2(n-1)/(4(n-1-J)(n-J))$.
\end{itemize}
To compare max$|S|$ between case 1 and case 2,
\begin{eqnarray}
(1-n_{\text{max}}^{-1})(c_1-c_0)^2/(n-J) &* & (c_1-c_0)^2(n-1)/(4(n-1-J)(n-J)) \notag\\
(1-n_{\text{max}}^{-1}) &*& (n-1)/(4(n-1-J)\notag\\
n_{\text{max}} &* &  4(n-1-J)/(3(n-1)-4J)   = \textstyle 1+(3-\frac{4J}{n-1})^{-1} \label{eqn:def2sp}
\end{eqnarray}
The right hand side (RHS) in Eq (\ref{eqn:def2sp}) $<$ the left hand side (LHS) $n_{\text{max}}$ except  when $n=2J$ (exactly 2 observations per cell in $\x$), which rarely happens in real-life data.  Therefore,
$$\textcolor{red}{\delta_{s_p}=
\begin{cases}
(c_1-c_0)^2 \frac{1-n_{\text{max}}^{-1}}{n-J} & \text{if $n>2J$}\\
(c_1-c_0)^2\frac{n-1}{n(n-2)} & \text{if $n=2J$ (exactly 2 observations per cell)}\\
\end{cases}},
$$
which is approximated by
$$\textcolor{red}{\delta_{s_p}=
\begin{cases}
(c_1-c_0)^2 (n-J)^{-1} & \text{if $n>2*J$}\\
(c_1-c_0)^2\frac{n-1}{n(n-2)} & \text{if $n=2*J$ (exactly 2 observations per cell)}\\
\end{cases}}\blacksquare
$$
if $n_{\text{max}}$ is large or when $n_{\text{max}}$ itself cannot be released without sanitization.

\section{\large{\texorpdfstring{$l_1$}{}-GS of pooled sample covariance}}\label{app:GScovp}
Denote the number of cells by $J$ and the number of observations in cell $j$ by $n_j$ ($j=1,\ldots,J$). Assume each cell contributes to the pooled covariance $\cov_{p}$; that is, each cells has at least 2 observations ($n_j\ge2$ for $j=1,\ldots,J$). Denote total sample size by $n=\sum_{j=1}^Jn_j$, then the pooled covariance between variables $x$ and $y$ is
$$\cov_{p}=\frac{\sum_{j=1}^J \sum_{i=1}^{n_j} (x_{ij}-\bar{x}_j)(y_{ij}-\bar{y}_j)}{n-J}$$

\textbf{Def 1}: WLOS, suppose it is the 1st observation in cell $j=1$ that differs between two data sets, then
\begin{align*}
&\Delta\!=\!\cov_p-\cov_p\!=\! (n-J)^{-1}\!\left(\!\textstyle\sum_{j=1}^J \sum_{i=1}^{n_j} (x_{ij}-\bar{x}_j)(y_{ij}-\bar{y}_j)\!-\!\textstyle\sum_{j=1}^J \sum_{i=1}^{n_j} (x'_{ij}-\bar{x}'_j)(y'_{ij}-\bar{y}'_j)\right)\\
&=(n-J)^{-1}(n_1-1) \underbrace{(n_1-1)^{-1}\left(\textstyle\sum_{i=1}^{n_1} (x_{i1}-\bar{x}_1)(y_{ij}-\bar{y}_j)-\sum_{i=1}^{n_1} (x_{i1}-\bar{x}'_1)(y'_{ij}-\bar{y}'_j)\right)}_{\text{term 1}}.
\end{align*}
Term 1 is the difference of the sample covariance in cell 1 between $\x$ and $\x'$, the  maximum of which is $n^{-1}_1(c_{11}-c_{10})(c_{21}-c_{20})$ per the results in Section \ref{app:GScov}. Therefore, max$|\Delta |\le(n-J)^{-1}\frac{(n_1-1)}{n_1}(c_{11}-c_{10})(c_{21}-c_{20})^2=(n-J)^{-1}(1-n^{-1}_1)(c_1-c_0)^2$, which again reaches the maximum if the $n_1$ is the largest among $(n_1,\ldots,n_J)$. All taken together,  the GS of $s_p^2$ is
\begin{align*}
\textcolor{red}{\delta_{s_p^2} = (c_{11}-c_{10})(c_{21}-c_{20})(n-J)^{-1}\left(1-n_{\text{max}}^{-1}\right)}, & \text{ which can be approximated by }\\
\textcolor{red}{\delta_{s_p^2} = (c_{11}-c_{10})(c_{21}-c_{20})(n-J)^{-1}}
\end{align*}
if $n_{\text{max}}$ is large or when $n_{\text{max}}$ itself cannot be released without sanitization.  $\quad \blacksquare$

\textbf{Def 2}:
WLOS, suppose it is the 1st observation in cell $j=1$ that is removed, then
$$\cov_{p-}=\frac{\sum_{j=2}^J \sum_{i=1}^{n_j} (x_{ij}-\bar{x}_j)(y_{ij}-\bar{y}_j)+ \sum_{i=2}^{n_1} (x_{i1}-\bar{x}_{1-})(y_{i1}-\bar{y}_{1-})}{n-1-J}$$
 \small
\begin{align*}
& \mbox{Let }  \Delta=\cov_p-\cov_{p-}=\frac{(n-1-J)\left(\sum_{j=2}^J\sum_{i=1}^{n_j} (x_{ij}-\bar{x}_j)(y_{ij}-\bar{y}_j)+\sum_{i=1}^{n_1} (x_{i1}-\bar{x}_1)(y_{i1}-\bar{y}_1)\right)}{(n-J)(n-1-J)}\\
&\quad\quad -\frac{(n-J)\left(\sum_{j=2}^J \sum_{i=1}^{n_j} (x_{ij}-\bar{x}_j)(y_{ij}-\bar{y}_j)+ \sum_{i=2}^{n_1} (x_{i1}-\bar{x}_{1-})(y_{i1}-\bar{y}_{1-})\right)}{(n-J)(n-1-J)}\\
&=\frac{\sum_{i=1}^{n_1} (x_{i1}-\bar{x}_1)(y_{i1}-\bar{y}_1)-\sum_{i=2}^{n_1} (x_{i1}-\bar{x}_{1-})(y_{i1}-\bar{y}_{1-})-\cov_{p-}}{(n-J)}\\
&=\frac{\displaystyle\sum_{i=1}^{n_1} x_{i1}y_{i1}-n_1^{-1}((n_1-1)\bar{x}_{1-}+x_{11})((n_1-1)\bar{y}_{1-}+y_{11})- \sum_{i=2}^{n_1}x_{i1}y_{i1}+(n_1-1)\bar{x}_{1-}\bar{y}_{1-}-\cov_{p-}}{(n-J)}\\
&=\frac{ (1-n_1^{-1})x_{11}y_{11}+(1-n_1^{-1})\left(\bar{x}_{1-}\bar{y}_{1-}-\bar{x}_{1-}x_{11}-\bar{y}_{1-}y_{11}\right)- \cov_{p-}}{(n-J)}\\
&= \frac{ (1-n_1^{-1})(x_{11}-\bar{x}_{1-})(y_{11}-\bar{y}_{1-})-\cov_{p-}}{(n-J)}\\
&= \underbrace{\frac{ (1-n_1^{-1})(x_{11}-\bar{x}_{1-})(y_{11}-\bar{y}_{1-})- \frac{\sum_{i=2}^{n_1}(x_{i1}-\bar{x}_{1-})(y_{i1}-\bar{y}_{1-})}{(n-1-J)}}{n-J}}_{\text{term 1}}-
 \underbrace{\frac{\sum_{j=2}^J \sum_{i=1}^{n_j} (x_{ij}-\bar{x}_j)(y_{ij}-\bar{y}_j)}{(n-1-J)(n-J)}}_{\text{term 2}}
\end{align*}
\normalsize
Term 1 is independent from term 2, meaning how term 1 changes does not the affect of the value of term 2. max$|\Delta|$ occurs when  1)  term 1 reaches the maximum and term 2 is at  its minimum, or 2) term 1 reaches its minimum and term 2 is at its maximum, whichever is larger gives max$\Delta$. The maximum and minimum of term 1 can be obtained as follows. First,  We transform $x$ and $y$ by $x=(c_{11}-c_{10})^{-1}(x-c_{10})$ and $y=(c_{21}-c_{20})^{-1}(y-c_{20})$.  The transformed variables range from $[0,1]$.  Let $k_1=\#\{x_{i1}=1\}, k_2=\#\{y_{i1}=1\}$ and $k_3=\sum_i x_{1i}y_{i1}$ for $i=2,\ldots,n_1$ in cell 1. Thus $\bar{x}=(n_1-1)^{-1}k_1$, and $\bar{y}=(n_1-1)^{-1}k_2$. It is easy to obtain that $k_3\in[\max(0, k_1+k_2-(n_1-1)), \min(k_1,k_2)]$. WLOS, assume $k_2\le k_1$, thus $0\le k_1\le n_1-1, 0\le k_2\le k_1, \max(0, k_1+k_2-(n_1-1))\le k_3\le k_2$. Therefore, the range of the values that $k_2$ depends on $k_1$, and that of $k_3$ depends on $k_1$ and $k_2$.
\begin{itemize}
\item If $(x_{11},y_{11})=(0,0)$, then term 1 is
\begin{align*}
&\frac{(1-n^{-1}_1)(n_1-1)^{-2}k_1k_2-\frac{k_3-(n_1-1)^{-1}k_1k_2}{n-1-J}}{(n-J)} \\
=&\frac{(n-J+n_1-1)k_1k_2}{n_1(n_1-1)(n-J)(n-1-J)}-\frac{k_3}{(n-J)(n-1-J)}
\end{align*}
$\Delta$ is linear in $k_1,k_2$ and $k_3$, thus the minimum/maximum occurs at corners of $(k_1,k_2,k_3)$.
\begin{itemize}
\item If $k_1=0$, then $k_2=0, k_3=0$, and term 1 is $0$
\item If $k_1=n_1-1$, then $k_2\in[0,n_1-1], k_3=k_2$, and the term 1 is  $\frac{k_2}{n_1(n-J)}$,  thus its maximum is $\frac{n_1-1}{n_1(n-J)} $ when $k_2=n_1-1$; and its minimum is 0 when $k_2=0$
\end{itemize}
\textcolor{red}{ Therefore, when $(x_{11},y_{11})=(c_{10},c_{20})$, then  term 1 is $\in(c_{11}-c_{10})(c_{21}-c_{20})\left[0, \frac{1-n_1^{-1}}{n-J}\right]$}
\item If $(x_{11},y_{11})=(0,1)$, then term 1 is
\begin{align*}
&\frac{-(1-n^{-1}_1)(n_1-1)^{-2}k_1(n_1-1-k_2)-\frac{k_3-(n_1-1)^{-1}k_1k_2}{n-1-J}}{(n-J)} \\
=&\frac{(n-J+n_1-1)k_1k_2}{n_1(n_1-1)(n-J)(n-1-J)}-\frac{k_3}{(n-J)(n-1-J)}-\frac{k_1}{n_1(n-J)}
\end{align*}
$\Delta$ is linear in $k_1,k_2$ and $k_3$, thus the minimum/maximum occurs at corners of $(k_1,k_2,k_3)$.
\begin{itemize}
\item If $k_1=0$, then $k_2=0, k_3=0$, and  term 1 is  $0$
\item If $k_1=n_1-1$, then $k_2\in[0,n_1-1], k_3=k_2$, and the term 1 is  $\frac{k_2}{n_1(n-J)}-\frac{n_1-1}{n_1(n-J)}$,  thus its maximum is 0 when $k_2=n_1-1$; and its minimum is $-\frac{n_1-1}{n_1(n-J)} $ when $k_2=0$
\end{itemize}
\textcolor{red}{ Therefore, when $(x_{11},y_{11})=(c_{11},c_{20})$, then term 1 is $(c_{11}-c_{10})(c_{21}-c_{20})\in\left[\frac{n_1^{-1}-1}{n-J},0\right]$}
\item If $(x_{11},y_{11})=(1,0)$, then
 \begin{align*}
&\frac{-(1-n^{-1}_1)(n_1-1)^{-2}k_2(n_1-1-k_1)-\frac{k_3-(n_1-1)^{-1}k_1k_2}{n-1-J}}{(n-J)} \\
=&\frac{(n-J+n_1-1)k_1k_2}{n_1(n_1-1)(n-J)(n-1-J)}-\frac{k_3}{(n-J)(n-1-J)}-\frac{k_2}{n_1(n-J)}
\end{align*}
$\Delta$ is linear in $k_1,k_2$ and $k_3$, thus the minimum/maximum occurs at corners of $(k_1,k_2,k_3)$.
\begin{itemize}
\item If $k_1=0$, then $k_2=0, k_3=0$, and  term 1 is  $0$
\item If $k_1=n_1-1$, then $k_2\in[0,n_1-1], k_3=k_2$, and the term 1 also becomes $0$ regardless of the value of $k_2$
\end{itemize}
\textcolor{red}{ Therefore, when $(x_{11},y_{11})=(c_{10},c_{21})$, then term 1 is 0}
\item If $(x_{11},y_{11})=(1,1)$, then
 \begin{align*}
&\frac{(1-n^{-1}_1)(n_1-1)^{-2}(n_1-1-k_1)(n_1-1-k_2)-\frac{k_3-(n_1-1)^{-1}k_1k_2}{n-1-J}}{(n-J)} \\
=&\frac{1-n^{-1}_1}{n-J}+\frac{(n-J+n_1-1)k_1k_2}{n_1(n_1-1)(n-J)(n-1-J)}-\frac{k_3}{(n-J)(n-1-J)}-\frac{k_1}{n_1(n-J)}-\frac{k_2}{n_1(n-J)}
\end{align*}
$S$ is linear in $k_1,k_2$ and $k_3$, thus the minimum/maximum occurs at corners of $(k_1,k_2,k_3)$.
\begin{itemize}
\item If $k_1=0$, then $k_2=0, k_3=0$, and  term 1 becomes $\frac{n_1-1}{n_1(n-J)}$
\item If $k_1=n_1-1$, then $k_2\in[0,n_1-1], k_3=k_2$, and term 1 becomes $0$ regardless of $k_2$
\end{itemize}
\textcolor{red}{Therefore, when $(x_{11},y_{11})=(c_{11},c_{21})$, then  term 1 is $\in(c_{11}-c_{10})(c_{21}-c_{20})\left[0, \frac{1-n_1^{-1}}{n-J}\right]$}
\end{itemize}
Taken together, term 1 $\in\left[-\frac{1-n_1^{-1}}{n-J}, \frac{1-n_1^{-1}}{n-J}\right]$.  For term 2, the numerator  $\sum_{j=2}^J \sum_{i=1}^{n_j} (x_{ij}-\bar{x}_j)(y_{ij}-\bar{y}_j)=\textstyle\sum_{j=2}^J (n_j-1) \cov_j$. Since $-\sum_{j=2}^J (n_j-1)\sqrt{\var{(x_j)}\var{(y_j)}}\le\textstyle\sum_{j=2}^J (n_j-1)\cov_j \le \sum_{j=2}^J (n_j-1)\sqrt{\var{(x_j)}\var{(y_j)}}$, so $ -(c_{11}-c_{10})(c_{21}-c_{20})\frac{n-n_1}{4} \le\sum_{j=2}^J (n_j-1)\cov_j\le(c_{11}-c_{10})(c_{21}-c_{20})\frac{n-n_1}{4}$. Terms 1 and 2 taken   together, $\Delta\in(c_{11}-c_{10})(c_{21}-c_{20})\times\left[-\!\!\left(\!\frac{n_1-1}{n_1(n-J)}\!+\!\frac{n-n_1}{4(n-J)(n-n_1-J+1)}\!\right), \frac{n_1-1}{n_1(n-J)}\!+\!\frac{n-n_1}{4(n-J)(n-1-J)}\!\right]$, and the maximum of $|\Delta|$ is
\begin{equation}\label{eq:def2covp}
(n-J)^{-1}(c_{11}-c_{10})(c_{21}-c_{20})\left(1-\frac{1}{n_1}+\frac{n-n_1}{4(n-1-J)}\right),
\end{equation}
which reaches the maximum at $n_1=2\sqrt{n-1-J}$, being plugging back in Eq (\ref{eq:def2covp}), we have
$$\textcolor{red}{\delta_{\mbox{cov}_p} =
\frac{(c_{11}-c_{10})(c_{21}-c_{20})}{n-J}\left(1+\frac{n}{4(n-1-J)}-\frac{1}{\sqrt{n-J-1}}\right)}$$
Of course, $n_1$ being exactly $=2\sqrt{n-1-J}$ is likely not to occur in real life since $2\sqrt{n-1-J}$ is most likely to be fractional, so $\delta_{\mbox{cov}_p}$ as given above is not an exact bound.

\section{\large{Additional Results from Simulation Study 2}}\label{app:simulation}
Figures S\ref{fig:sim1sum} to S\ref{fig:sim1sumUH} on pages 11-12 of this document present the simulation results regarding the inferences of 
some linear combinations of $\mathbf{p}$ based on the synthetic data from the rescaling and the universal histogram approaches. A brief discussion on the results are presented in the main manuscript.

\section*{Reference}
Shiffler, R. E. and Harsha, P. D. (1980), Upper and lower bounds for the sample standard deviation, Teaching Statistics, 2(3):84-86

Hay, M.,  Rastogiz, V.,  Miklauy, G.  and Suciu, D. (2010) Boosting the Accuracy of Differentially Private Histograms Through Consistency, Proceedings of the VLDB Endowment, 3(1): 1021-1032

\begin{landscape}
\begin{figure}[H]
\centering \includegraphics[scale=0.75]{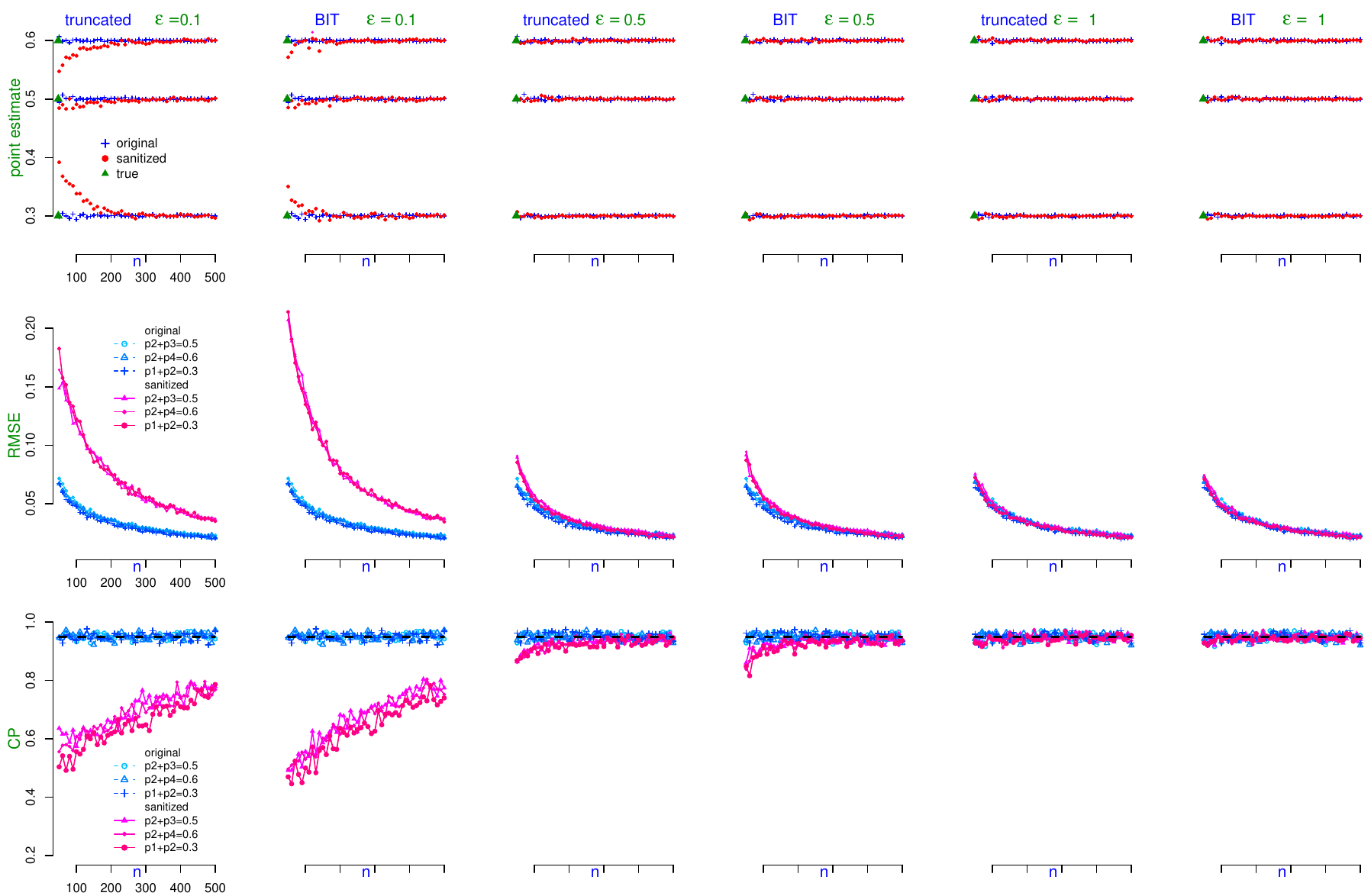}
\caption{Bias, RMSE and CP of sanitized proportions in the rescaling approach (red lines represent the original linear combinations ($p_1+p_2,p_1+p_3,p_1+p_4$) of $\mathbf{p}$, and blue lines represent the sanitized versions)}\label{fig:sim1sum}
\end{figure}
\begin{figure}[H]
\centering \includegraphics[scale=0.75]{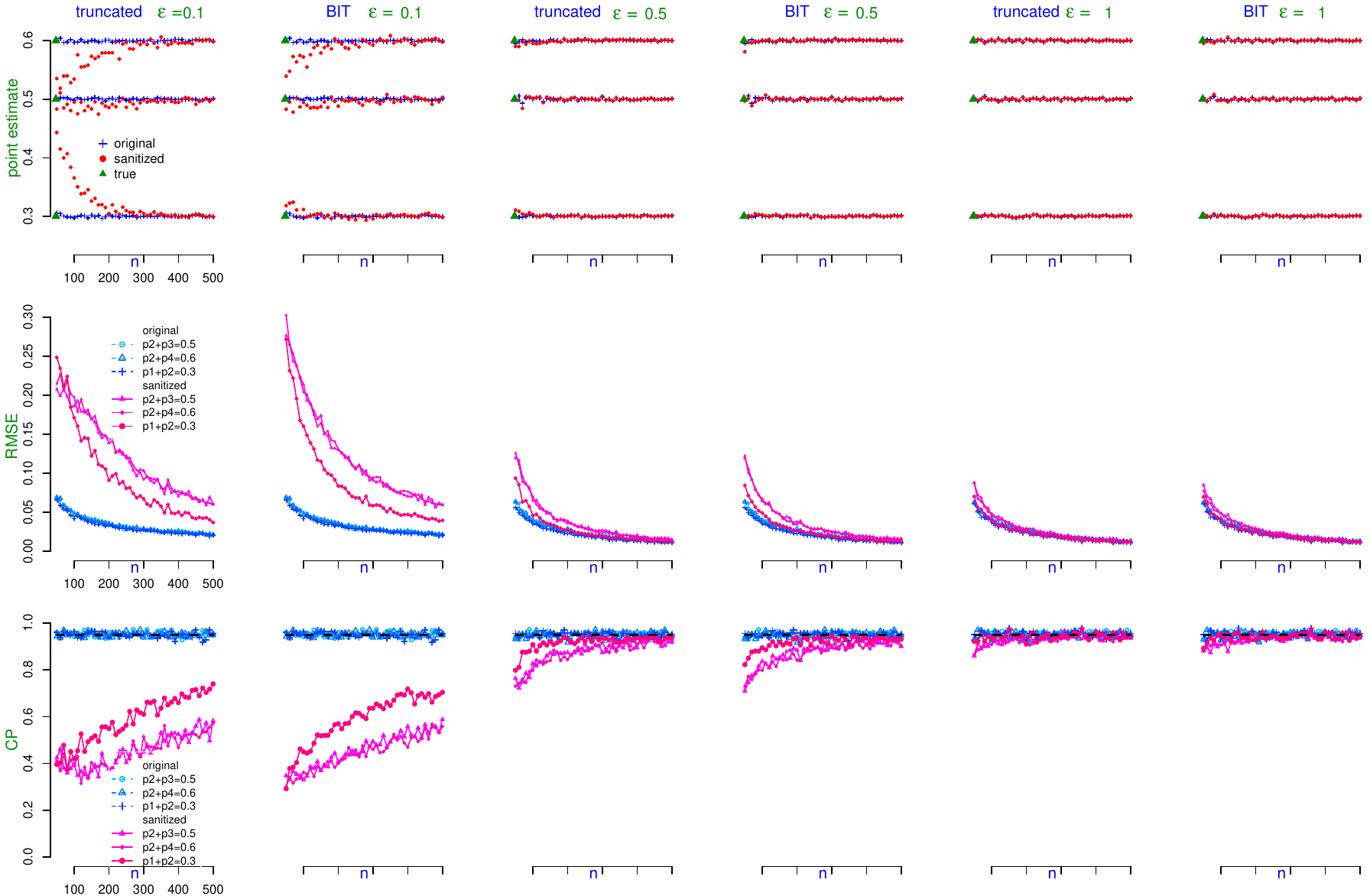}
\caption{Bias, RMSE and CP of sanitized proportions in a modified universal histogram procedure based on Hay et. al. (2010) (red lines represent the original linear combinations ($p_1+p_2,p_1+p_3,p_1+p_4$) of $\mathbf{p}$, and blue lines represent the sanitized versions)}\label{fig:sim1sumUH}
\end{figure}
 \end{landscape}
\end{document}